\documentclass[dvips]{acta}
\usepackage{supertabular,lscape,epsfig}
\usepackage{correct}
\usepackage{amssymb}
\usepackage{amsmath}
\usepackage{physics}
\usepackage{natbib}
\usepackage{graphicx}
\usepackage{verbatim}
\usepackage{subcaption}
\usepackage{float}
\usepackage{changepage}
\usepackage[linktoc=all]{hyperref}
\usepackage{appendix}

\hypersetup{
    breaklinks=true,   % splits links across lines
    colorlinks=true,   % displays links as colored text instead of blocks
    pdfusetitle=true,  % \title and \author values into pdf metadata
    implicit=false,       % etc.
 }

\SetPages{0}{0}

\SetVol{0}{2023}

\begin{document}

\begin{Titlepage}
\Title{Search for dormant black holes in the OGLE data}

\Author{M.~K~a~p~u~s~t~a$^1$,  P.~M~r~\char0243~z$^2$}
{ Astronomical Observatory, University of Warsaw, Al. Ujazdowskie 4, 00-478 Warsaw, Poland\\
    $^1$email: mr.kapusta@student.uw.edu.pl\\
    $^2$email: pmroz@astrouw.edu.pl
}

\Received{XXX-YYY-ZZZ}
\end{Titlepage}
\Abstract{Light curves of ellipsoidal variables collected by the Optical Gravitational Lensing Experiment (OGLE) were analyzed, in order to  
search for dormant black hole candidates. After the preselection based on the amplitude of ellipsoidal modulation, each object was investigated by means
of the spectral energy distribution fit, which allowed us to select objects that are in close agreement with the spectrum of a single stellar object. 
After this final step of the preselection process, we were left with only fourteen objects that were then investigated in detail.
For each candidate, we estimated basic physical parameters such as temperature, mass, luminosity, and, in some cases, radial velocity semi-amplitude.
One of the objects turned out to be a spotted star while the rest are considered black-hole binary candidates.
In the end, we present an alternative explanation for the ellipsoidal modulation in the form of contact binaries, which are not only vast in number, contrary 
to black-hole binaries, but are also in much better agreement with the radial velocity estimates for some of the systems analyzed. 
Even if the presented arguments suggest a noncompact character
of the companion objects, each of them should be observed spectroscopically in order to verify the compact object hypothesis.
}{compact objects, ellipsoidal variable stars, black holes, neutron stars, binary star systems}

\section{Introduction}
According to current knowledge, O type stars (which are more massive than $20$\,M$_\odot$) should end their evolution as black holes. Further investigation indicates
the existence of $\mathcal{O}(10^8)$ stellar-mass black holes \citep{olejak_catalog} in our Galaxy.
Despite a huge theoretical expectation, only a few tens of such objects have been found up to this day in the Milky Way, all but one in binary systems. 
Most of those black holes have been found due to an X-ray emission caused by the accretion disk in the system. 
Unfortunately, not only the emission from such systems is transient, but also various estimates suggest that only a tiny fraction of them can be found in
close, accreting binaries \citep{portegies_zwart_formation_1997}. It is estimated that from the whole black-hole binary population only around $\sim 1000$ objects are expected to produce
X-ray signal \citep{corral-santana_blackcat_2016}. In wider systems the accretion rate is too low to produce a detectable X-ray emission. 

In 2016 a new observational window was opened, as the LIGO/Virgo collaboration announced the detection of the first black hole-black hole merger \citep{LIGO_first}.
In subsequent observing runs, more detailed measurements of masses of merging black holes and neutron stars in dozens of events have been made \citep{LIGO_catalog}, broadening our knowledge about
the formation mechanisms of black holes. 
Such observations allow us not only to test current evolutionary models, but also
set expectations for observational verification of formation channels.
Observation of a more representative population would certainly allow better
insight into long-standing problems such as the formation of the mass gap objects observed by the LIGO/Virgo.

In recent years new methods have been developed with the goal of identifying black-hole binaries that do not
emit X-ray radiation, called dormant black holes. Up to the day, multiple candidates have been identified using spectroscopic observations \citep{shenar_x-ray-quiet_2022},
gravitational microlensing \citep{sahu_isolated_2022,lam_isolated_2022,mroz_systematic_2022}, and astrometric observations
\citep{el-badry_sun-like_2022,el-badry_red_2023}. Despite those advances, the quest for dormant black holes is still ongoing
and next generations of sky surveys are expected to produce new potential candidates. Both Gaia astrometric observations and
spectroscopic observations are limited, contrary to photometric observations that are vast in number thanks to photometric surveys
such as OGLE \citep{udalski_optical_1992}, ZTF \citep{ztf_ref}, TESS \citep{TESS_ref}, or soon-to-start LSST \citep{LSST_ref}. Hence, if black holes could be successfully
detected using photometric data only, they could achieve a much greater potential yield compared to spectroscopic surveys.

Motivated by the recent advances in the field, a new method was introduced with the aim of uncovering potential black-holes resting in dormant systems based on its ellipsoidal
modulation \citep{gomel_search_2021a,gomel_search_2021b,gomel_search_2021c}.
The relative potential of a similar method was previously evaluated by \citet{masuda_prospects_2019}. It was found that at least a few dozen of them should
be detectable in the TESS survey data, based on the relativistic beaming effect paired with the ellipsoidal modulation.

Here, an application of the aforementioned method to the OGLE data is presented. In Section 2, a general theoretical introduction is given. Next, in Section 3,
a detailed two-step selection procedure is presented.
In Section 4, detailed analyzes of the objects are provided, along with an alternative explanation for the ellipsoidal modulation. Section 5 contains a discussion and conclusions.

\section{Selection method}
In this work, the method based on the original approach presented in \citet{gomel_search_2021a,gomel_search_2021b,gomel_search_2021c} was adopted. 
The authors presented how to obtain a lower bound on the mass ratio using the second harmonic coefficient of the light curve dominated by the
ellipsoidal modulation.
They found that, based on a mass ratio $q$, an orbital inclination $i$, a fillout factor $f$, and a particular parameter $\alpha_2$, the second harmonic coefficient
can be calculated by the equation
\begin{equation}
    A_2(q,f,i,\alpha_2)=\frac{C_2(q,f,i)}{1+\frac{1}{9}E(q)^3(2+5q)(2-3\sin^2{i})} \alpha_2 E(q)^3 f^3 q \sin^2{i}.
\end{equation}
$C_2$ is a correction function introduced by the authors, while $E(q)$ stands for the ratio of the Roche lobe radius to the semi-major axis obtained using the
Eggleton formula \citep{eggleton_approximations_1983}. As the amplitude is an increasing function of $f$ and $\sin{i}$, the authors introduced the \textit{ modified mass ratio} (denoted 
as $q_{\textrm{mmr}}$) that acts as a lower bound for the true mass ratio. It can be obtained after solving the equation
\begin{equation}
    A_2(q_{\textrm{mmr}},0.98,\pi/2,1.2) = A_2,
\end{equation}
where $A_2$ is a measured value of the second harmonic coefficient. 
$\alpha_2$ parameter depends on the limb and gravity darkening coefficients, which are dependent on temperature, metallicity, etc.
Its value in the equation was assumed to be equal to $1.2$
(for measurements in the $I$ band) following the prescription of \citet{gomel_search_2021a,gomel_search_2021b,gomel_search_2021c}.
According to the authors, the value of $1.2$ should be typical for main sequence stars although it varies with the standard deviation of $0.1$ depending on the temperature,
prompting the authors to introduce a lower bound of $q_{\textrm{mmr}}$.
The second parameter denoted as $\tilde{q}_{\textrm{mmr}}$ is defined using the following equation:
\begin{equation}
    A_2(\tilde{q}_{\textrm{mmr}},0.98,\pi/2,1.3) = A_2 - \Delta A_2,
\end{equation} where $\Delta A_2$ denotes the uncertainty associated with the estimation of $A_2$. This is a slightly modified version of the original approach, in which not only was the 
relative uncertainty of $\alpha_2$ taken into account (by increasing $\alpha_2$ in the definition to $1.3$), but also light curves with high uncertainties associated with the estimate $A_2$
were penalized. Therefore, in the end, the relationship 
\begin{equation}
    q>q_{\textrm{mmr}}>\tilde{q}_{\textrm{mmr}}
\end{equation}
should hold.
Following the original approach presented in \citet{gomel_search_2021a,gomel_search_2021b,gomel_search_2021c}, objects with $\tilde{q}_{\textrm{mmr}}>1$ are selected to the next stage. 
The exact threshold can be a matter of debate; in \citet{gomel_gaia_2022} the threshold of $0.5$ was selected as theoretically the true mass ratio should be much higher
than the estimate of the lower bound. Here, the original threshold was used to impose a stronger cut on the sample that would be subjected to a further selection process. 

\section{Observational data}
\subsection{Description of data and preprocessing}
\begin{figure}
    \begin{center}
        \includegraphics[width = \textwidth]{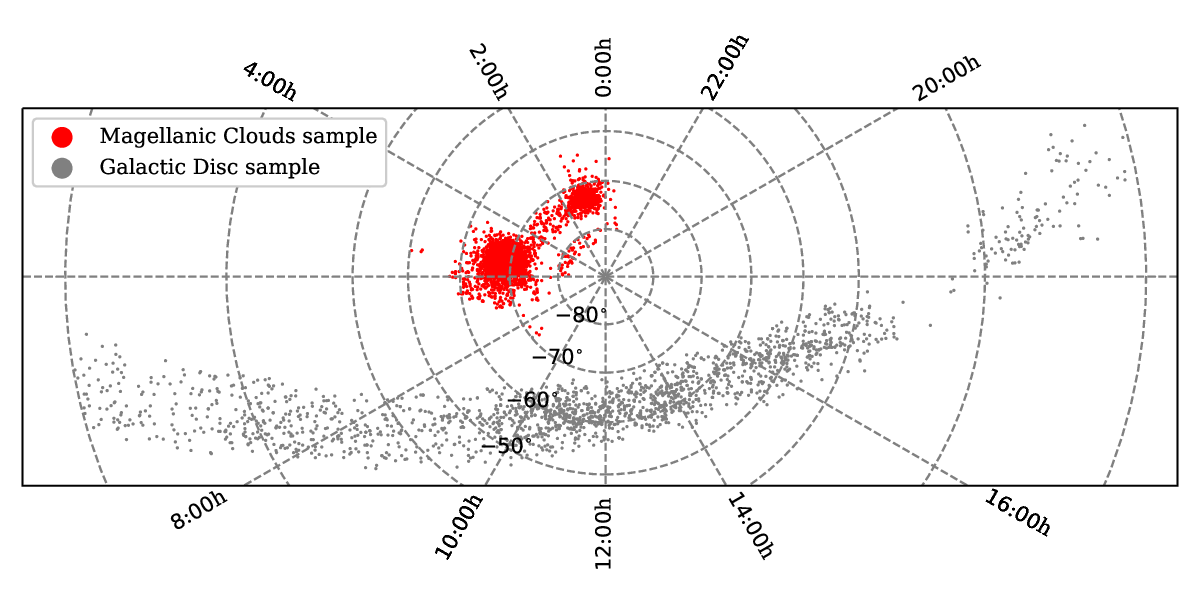}
    \end{center}
    \caption{Positions of objects from the samples across the southern sky.}
    \label{map}
\end{figure}

Light curves used in the analysis were collected during the fourth phase of the OGLE \citep{udalski_ogle-iv_2015} survey in the $I$ band.
Objects are divided into two separate samples: Magellanic Clouds and the Galactic plane.
The first sample is based on the stars presented in \citet{pawlak_ogle_2016} with additional objects supplemented.
The second sample consists of the ellipsoidal stars that were cross-matched with the Gaia DR3 \citep{gaia_collaboration_gaia_2023}; objects with 
\textit{rv\_amplitude\_robust} value (radial velocity amplitude after outlier removal) greater than $100$ km/s were selected.
The first sample can be divided into objects from the direction toward the Large Magellanic Cloud (LMC), the Small Magellanic Cloud (SMC), and the Magellanic Bridge (MBR).
The second sample is divided into two parts, objects from the direction toward the Galactic bulge (BLG), and the Galactic disk (DG or GD depending on whether an object is west/east of the Galactic bulge,
respectively). Detailed positions of objects from the original samples in the sky are presented in Figure \ref{map}.

\begin{figure}
    \begin{center}
        \includegraphics[width = \textwidth]{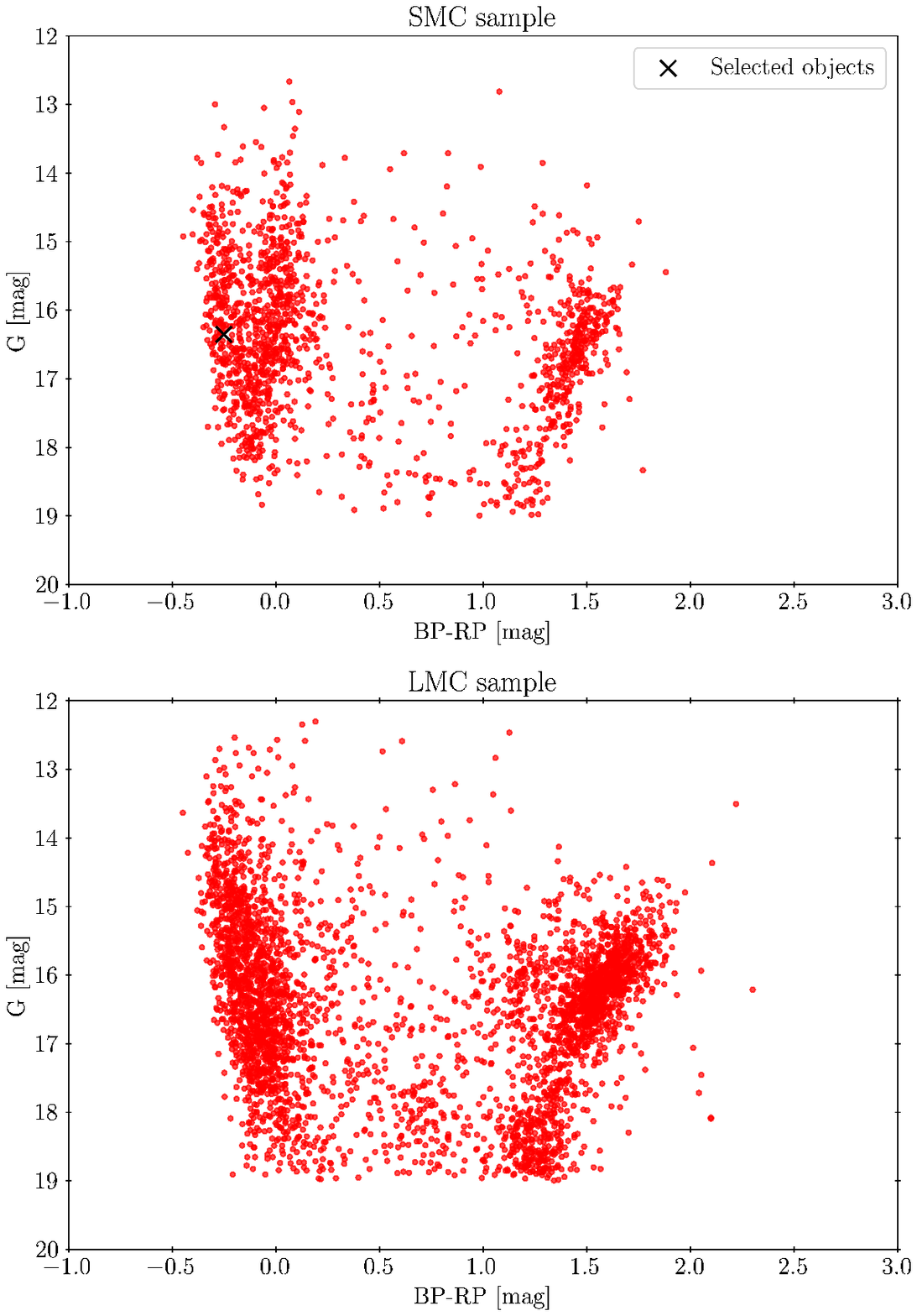}
    \end{center}
    \caption{A color-magnitude diagram of Gaia counterparts potentially residing in Magellanic Clouds (without statistically significant parallax).}
    \label{HR_SMC}
\end{figure}

\begin{figure}
    \begin{center}
        \includegraphics[width = \textwidth]{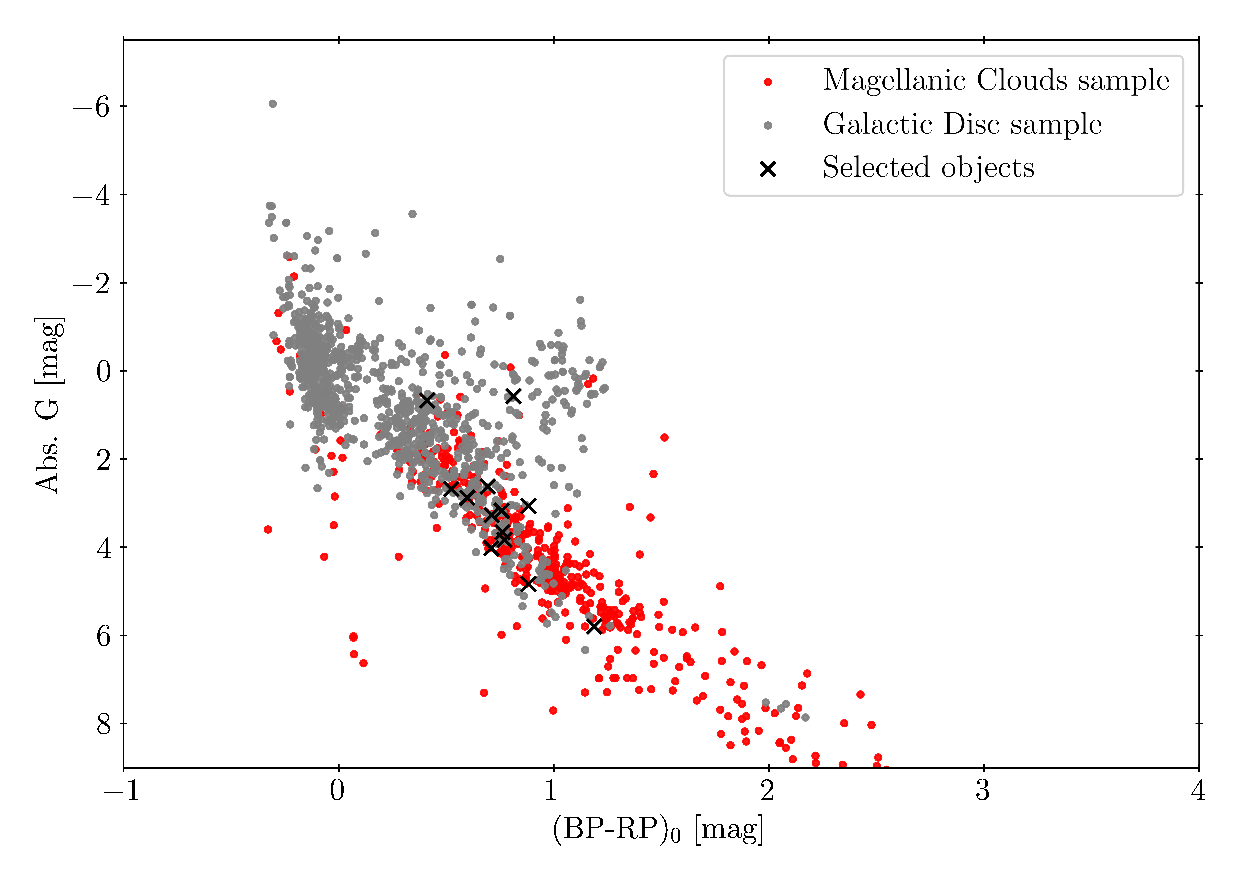}
    \end{center}
    \caption{A color-magnitude diagram of Gaia counterparts with statistically significant parallax.}
    \label{HR_galactic}
\end{figure}

\begin{figure}
    \includegraphics[width = \textwidth]{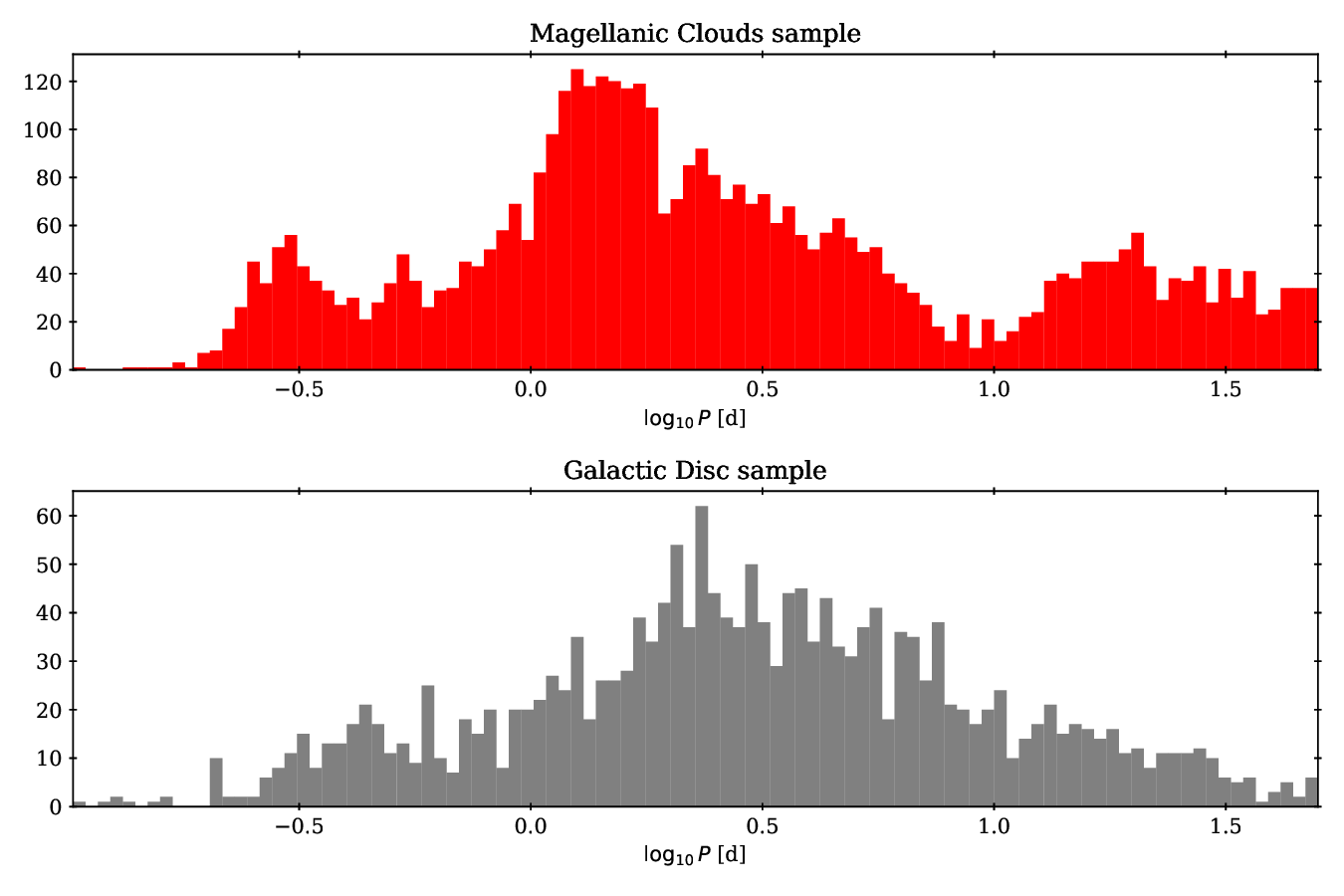}
    \caption{Distribution of periods from the analyzed samples.}
    \label{periods}
\end{figure}

In general, objects from the direction toward the Magellanic Clouds can include stars that 
only accidentally line up. To determine the membership of the stars (Magellanic Clouds or accidental alignment),
each entry has been cross-matched with Gaia DR3 catalog. Objects with statistically insignificant parallaxes or objects with wrong astrometric solutions (RUWE $>1.4$)
were assumed to reside in the Magellanic Clouds, while the rest were considered to belong to the Galactic population. Based on the memberships, two color-magnitude diagrams 
have been constructed. Figure \ref{HR_SMC} presents stars assumed to reside in the Magellanic Clouds, with the division between SMC and LMC.
Figure \ref{HR_galactic} presents color-magnitude diagram for the rest of the stars.
While in the first figure, the directly measured values were plotted, in the second the absolute values of the brightness in the $G$ band were calculated (corrected for the extinction 
based on the $A_G$ values processed by the Gaia pipeline) and presented on the y axis.
Colors of the objects were also corrected using the E(BP-RP) estimates given by Gaia. 
On top of the diagrams objects selected at the end of this section were overplotted with the black cross.  
Most of objects in the samples have a relatively short period ($<3$ d); a detailed period distribution
is presented in Figure \ref{periods}.

As a first step of preprocessing, objects fainter than
$17$ mag were removed, as faint stars would not be suitable for a radial velocity determination high-resolution spectroscopy.
Each object was reanalyzed using the AOV (analysis of variance) method \citep{schwarzenberg-czerny_advantage_1989} to refine the period estimation (denoted hereafter as $P$).
Uncertainties associated with brightness measurements were recalculated using the method presented in \citet{uncertainties_OGLE}.
Then, each light curve (denoted as $I(t)$) was fitted with the $4$th degree harmonic model
\begin{equation}\label{harm}
    I(t)=A_0+\sum_{i=1}^4\left[ A_{1i}\sin{\left(2\pi i\frac{t}{P}\right)}+A_{2i}\cos{\left(2\pi i\frac{t}{P}\right)}\right]
\end{equation}
with the sigma clipping threshold set at $3\sigma$, allowing us to determine amplitudes of coefficients from the relationship $A_i=\sqrt{A_{1i}^2+A_{2i}^2}$.
This least-squares problem was solved using the Levenberg-Marquardt algorithm with an error estimate based on a covariance matrix.
Subsequently, the minimum mass ratio $q_{\textrm{mmr}}$ and its lower bound $\tilde{q}_{\textrm{mmr}}$ were calculated and the previously described procedure for selecting candidates was carried out.
Objects with light curves indicating another type of variability than ellipsoidal one were removed together with those with periods longer than
$50$ d. According to the main assumption in the analysis, objects should be composed of stars nearly filling their Roche lobe, so a rather short period is suggested.
The exact value of the threshold can be debated; here it is used as a means to remove pulsating stars that pollute the sample.
After this part of the preprocessing, only $41$ objects from the first sample were left, together with $22$ objects from the second sample.

\subsection{Spectral Energy Distribution}
Each object was cross-matched against the Gaia DR3 catalogue to obtain parallax (hereinafter denoted as $\pi_0$ together with its uncertainty denoted as $\sigma_{\pi}$)
estimates. If the parallax was statistically significant ($\pi>3\sigma_{\pi}$)
it was used to derive the distance to the object. For objects toward Magellanic Clouds, many entries lacked a statistically significant parallax, indicating that they indeed reside
in the Magellanic Clouds.
In the case of LMC/MBR and SMC, the assumed distance estimates from \citet{jacyszyn-dobrzeniecka_ogle-ing_2016} 
were $d_{\textrm{LMC}}=49.93\pm1.79$ kpc and $d_{\textrm{SMC}}=64.62\pm4.95$ kpc, respectively.

To further investigate the nature of objects, a spectral energy distribution analysis was
performed with two models: a single star model and a double star model. The main goal of this type of analysis is to use photometry from various bands to
reconstruct a whole spectral characteristic and hence infer parameters (such as temperature and luminosity) of an object.
The single star model depends on three free parameters: a logarithm of temperature $\log{T}$ (expressed in Kelvins), a logarithm of luminosity $\log{L}$
(expressed in $L_{\odot}$) and the third parameter is either a parallax (if one was measured) or a distance. In the case of the double-star model,
there were two additional parameters describing the second star $\log T_2$ and $\log L_2$. In the case of the first model, 
the star was assumed to be a main sequence object (MS) with $\log{g} = 4$.
In the double star model, the primary star was assumed to have $\log{g_1}=4$ as in the first case, while the second object was assumed to be a giant with $\log{g_2}=2$. 
Objects with a statistically significant parallax were assumed to have solar-like
metallicity ($Z = 0.013$) as they likely reside inside the Galactic disk. Objects in the Magellanic Clouds
were assumed to have a metallicity equal to $Z = 0.010$ in the case of the LMC/MBR and $Z = 0.005$ in the case of the SMC.
The BaSeL library of stellar spectra \citep{lejeune_standard_1998} was used to find the spectrum for any given $\log{T}$, $\log L$, $d$ 
by means of the interpolation as implemented in the Python library
\texttt{pystellibs}\footnote{https://github.com/mfouesneau/pystellibs}.
Moreover, as some stars reside in high extinction regions, B-V color excess is used to redden the spectrum according to some extinction law.
The value of $E(B-V)$ is treated similarly to $Z$/$\log g$ as the hyperparameter and was estimated from the external dust maps.
In the calculations, the Cardelli extinction law \citep{cardelli_relationship_1989} with $R_V=3.1$ was assumed
and the Python implementation from \texttt{extinction}\footnote{https://extinction.readthedocs.io/en/latest/index.html} was used. 
Then the theoretically calculated stellar spectrum was processed using the \texttt{pyphot}\footnote{https://github.com/mfouesneau/pyphot} 
Python library, yielding a magnitude in a required filter.

In order to find a set of best-fitting parameters, a Bayesian approach was adopted. One can denote a set of observed fluxes as $\tilde{f}_i$, theoretically predicted
fluxes as $f_i$, and errors of fluxes as $\sigma_i$. Each of the measured fluxes $\tilde{f}_i$ is expected to be normally distributed with a mean $f_i$ and a standard deviation $\sigma_i$,
while the prior distributions for each of the three
parameters are uniform in the case of $\log T$ and $\log L$, or normal in the case of a distance or parallax.

As the sources are selected due to the high amplitude of ellipsoidal modulation, flux measurements are subjected to further deviations from the mean, as some surveys 
could obtain measurements of the source at different phases. In order to take into account this effect, it was assumed that the relative change of flux does not depend on the band. 
This type of approximation is true in the blackbody limit as the change of the flux depends only on the apparent surface. Hence standard deviation in $i$th filter can be obtained as 
\begin{equation}
    \sigma_i = \sqrt{(0.4 \ln{10}\tilde{f}_i\sigma_{m})^2+\sigma_{\textrm{inst}^2}},
\end{equation}
where $\sigma_{\textrm{inst}}$ is the uncertainty tied to the flux measurement while $\sigma_{m}$ stands for the standard deviation of the magnitude measurements in the $I$ band
reported by OGLE. Similar treatment was applied to the double star model. 
We would like to split total calculated flux in the $i$th filter into two contributing fluxes corresponding to stars denoted respectively as 
$f_{1i}$ and $f_{2i}$. 
After similar calculations we arrive at 
\begin{equation}
    \sigma_i =\sqrt{\left(0.4 \ln{10}\tilde{f}_{1i}\sigma_{m} \frac{1}{R}\right)^2+\sigma_{\textrm{inst}^2}},
\end{equation} where 
\begin{equation}
    R(\log T_1,\log L_1,\log T_2,\log L_2)=\frac{f_{1\textrm{I}}}{f_{1\textrm{I}}+f_{2\textrm{I}}}.
\end{equation} Here the first component is assumed to be variable and subscript I stands for the predicted value in the $I$ band.
If the first component is much more luminous in the $I$ band, we approximate $R\approx 1$ obtaining an expression similar to the error correction from single star model.
As this expression is not symmetrical, there is a need to determine which component is variable. Here the following procedure was adopted.
First, the initial sampling was performed with instrumental errors to estimate the values in the system. Then, the more luminous component in the $I$ band was considered to be variable 
and the second run was performed with estimated error based on the ellipsoidal modulation. This treatment can lead to an overestimation of the error as some surveys (like Gaia)
use many measurements to estimate the brightness, effectively reducing the expected scatter of the measurements. Even if both models are corrected for the ellipsoidal modulation,
one can expect a slight bias toward the single star model, as expected errors are inherently lower than in the double model case. Although it might affect judgment in the case where one
of the stars is rather faint, it should be noted that if both stars contribute to the spectrum significantly, no single star model can explain photometry of the system 
even after error correction.

A logarithm of luminosity was set in the boundary $(-3,5)$ to eliminate possible unphysical solutions, while the boundary on a logarithm of temperature
was based on the boundaries of the stellar library (which depend on surface gravity).
Under those assumptions, a log-likelihood function can be written as 
\begin{align}
    \begin{split}
    \ln{\mathcal{L}} &= -\sum_i\frac{(\tilde{f}_i-f_i)^2}{2\sigma_i^2}-\frac{(d-d_0)^2}{2\sigma_d^2} + C_1\textrm{ or }\\
    \ln{\mathcal{L}} &= -\sum_i\frac{(\tilde{f}_i-f_i)^2}{2\sigma_i^2}-\frac{(\pi-\pi_0)^2}{2\sigma_{\pi}^2} + C_2,
    \end{split}
\end{align} depending on whether a distance or a parallax was used, where $C_1$ and $C_2$ are constants. 

The following catalogs were used to assemble SEDs:
\begin{enumerate}
\item Catalogs shared by both samples of objects:
\begin{itemize}
    \item 2MASS survey \citep{skrutskie_two_2006},
    \item Gaia DR2 \citep{gaia_collaboration_gaia_2018},
    \item VISTA Hemisphere Survey DR5 \citep{mcmahon_vizier_2021},
    \item ALLWISE/WISE survey \citep{wright_wide-field_2010,cutri_vizier_2021},
    \item GALEX Survey \citep{bianchi_galex_2011},
    \item Denis survey  \citep{denis_vizier_2005},
    \item SkyMapper DR1/DR2 \citep{wolf_skymapper_2018,onken_skymapper_2019},
    \item XMM Optical Monitor serendipitous sources catalog \citep{page_xmm-newton_2012}.
\end{itemize}
\item Catalogs exclusive to the Magellanic Clouds sample:
\begin{itemize}
    \item VISTA Magellanic Cloud survey DR4 \citep{cioni_vizier_2017},
    \item Spitzer SAGE survey: SMC and LMC \citep{meixner_spitzer_2006,gordon_surveying_2011},
    \item Denis catalog of objects in Magellanic Clouds \citep{cioni_denis_2000},
    \item Magellanic Clouds Photometric Survey: SMC and LMC \citep{zaritsky_magellanic_2002,zaritsky_magellanic_2004}.
\end{itemize}
\item Catalogs exclusive to the Galactic disk sample:
\begin{itemize}
    \item Bochum Galactic disk survey \citep{hackstein_bochum_2015},
    \item AAVSO Photometric All Sky Survey DR9 \citep{henden_apass_2015},
    \item VISTA Variables in Via Lactea Survey DR2 \citep{minniti_vizier_2017},
    \item GLIMPSE source catalog \citep{spitzer_science_vizier_2009}.
\end{itemize} 
\end{enumerate}
In the case of Magellanic Clouds, extinction estimates were based on the map of \citet{skowron_ogle-ing_2021}, while in the case of the Galactic disk, extinction
was obtained using the Python library \texttt{mwdust} \citep{bovy_galactic_2016} with the 3D dust map being a combination of \citet{green_3d_2019}, \citet{marshall_modelling_2006},
\citet{drimmel_three-dimensional_2003}.
As the OGLE dust map does not cover the positions of all objects, estimates for those were also obtained using the \texttt{mwdust} library.
Using the described setup, the Python-based library MCMC \texttt{ emcee}\footnote{https://emcee.readthedocs.io/en/stable/} \citep{foreman-mackey_emcee_2013}
was used to construct the set\footnote{https://github.com/Wesenheit/Iris} of routines used to sample from the posterior of the models, providing estimates of parameters together
with associated uncertainties.
Other packages used in the study are \texttt{astroquery} \citep{ginsburg_astroquery_2019},
\texttt{corner} \citep{foreman-mackey_cornerpy_2016} and \texttt{astropy} \citep{astropy_collaboration_astropy_2022}.
Each sampling was carried out with $32$ walkers for $6000$ steps (with $2250$ step burn-in); in the case of a double model, each sampling 
was followed by a second one with starting conditions sampled from the first chain. 

In order to help decide between models, the BIC score was used
\begin{equation}
    BIC=k\log{n}-2\ln{\mathcal{L}},
\end{equation} where $k$ is a number of estimated parameters and $n$ is a number of data points. Out of all objects, thirteen were selected based on the
fit with the single star model.
If the BIC score was lower in the case of the single star model, then the star was selected as a candidate.
If the BIC score was lower in the case of the double star model, then the object was rejected unless
\begin{itemize}
    \item the most luminous star in the double star model was significantly brighter than the second one and had $\log g=2$,
    \item some stars in the model had unphysical solutions (i.e. luminosity was too low for a star with $T=20000$ K).
\end{itemize}
There might be a possibility that the photometry can be better fitted with lower gravity, and hence, even though the object
is single in nature, it will always be fitted better with a double star model (because in a single star model $\log g=4$ while in a double star model
one star has $\log g=2$).
There can be also a possibility, that a star with a slight excess in UV will be better fitted in the double star model than with the single star model
because the metallicity in the single star model is slightly wrong. All of those factors are taken into account
and each object was manually inspected to check if it qualifies as a single star. Moreover it should be noted that assuming the wrong values of hyperparameters can 
lead to the underestimation of the errors on parameters like temperature. Error smaller than $100$ K on the temperature should be treated as unphysical.

\begin{table}
    \scriptsize
    \centerline{
    \begin{tabular}{llllllllllll}
    \hline
    Name&     Period [d]&     $q_{\textrm{mmr}}$ & $\tilde{q}_{\textrm{mmr}}$   &    Ra [deg]        &  Dec [deg]        &   $A_0$    &      $A_1$ &   $A_2$     &    $A_3$     &     $A_4$  & $\pi_0$ [mas] \\
    \hline
    \hline
    BLG931.27.36745 & $0.705044$   & $2.44$  & $1.63$ & $261.74489$  & $-40.31792$  & $11.6331$ & $0.0033$ & $0.1113$ & $0.0026$ & $0.0105$ & $1.456$     \\[0.1cm]
    BLG986.08.7     & $0.513215$  & $1.46$ & $1.03$  & $260.451241$ & $-43.019464$ & $12.1921$ & $0.0028$  & $0.0985$ & $0.0039$  & $0.0093$ & $1.199$ \\[0.1cm]
    GD1070.18.22288 & $45.146726$  & $7.41$ & $3.73$ & $257.007546$ & $-41.048747$ & $11.3696$ & $0.0091$  & $0.1362$  & $0.0043$  & $0.0049$ &   $ 1.031$ \\[0.1cm]
    GD1097.20.23000 & $0.457069$ & $9.73$ & $5.18$  & $252.108419$ & $-44.137004$ & $11.7822$ & $0.0008$ & $0.1414$  & $0.0029$ & $0.0200$  &  $1.775$ \\[0.1cm]
    GD1448.27.17    & $1.241002$  & $1.77$ & $1.23$ & $139.969949$ & $-45.759178$ & $11.8754$ & $0.0115$  & $0.1035$ & $0.0030$  & $0.0056$  &  $ 0.769$ \\[0.1cm]
    GD2246.03.18414 & $0.427085$ & $2.29$ & $1.55$ & $179.327372$ & $-57.091968$ & $12.0765$ & $0.0045$  & $0.1090$ & $0.0030$ & $0.0106$  & $1.410$ \\[0.1cm]
    LMC574.11.3407  & $0.255128$ & $1.48$ & $1.03$ & $80.1415$    & $-63.185667$ & $16.9192$ & $0.0064$ & $0.0989$ & $0.0013$ & $0.0092$  &  $ 0.454$ \\[0.1cm]
    LMC606.30.48    & $0.269887$   & $1.67$  & $1.11$ & $92.27575$   & $-63.376167$ & $15.9066$ & $0.0141$  & $0.1019$ & $0.0038$  & $0.0122$ &   $ 0.518$ \\[0.1cm]
    LMC751.15.2886  & $0.375382$ & $1.70$ & $1.16$  & $98.082958$  & $-66.307861$ & $16.9589$    & $0.0090$  & $0.1024$ & $0.0012$ & $0.0121$  &  $ 0.182$ \\[0.1cm]
    MBR108.18.3     & $0.294733$  & $1.49$ & $1.02$ & $33.257292$  & $-72.993167$ & $13.3766$ & $0.0063$ & $0.0989$ & $0.0033$  & $0.0128$  &  $ 1.097$ \\[0.1cm]
    MBR236.09.433   & $0.439879$  & $1.51$ & $1.07$ & $50.642958$  & $-80.540667$ & $14.5856$ & $0.0024$ & $0.0994$  & $0.0028$ & $0.0089$  &  $ 0.473$ \\[0.1cm]
    SMC711.22.1068  & $0.446659$  & $2.59$  & $1.73$ & $9.833875$   & $-70.378028$ & $13.2263$ & $0.0026$  & $0.1128$ & $0.0056$  & $0.0167$ & $ 0.660$     \\[0.1cm]
    SMC720.28.40576 & $0.567448$ & $7.07$  & $4.09$ & $11.938042$  & $-73.13625$  & $16.3672$  & $0.0017$ & $0.1352$ & $0.0056$  & $0.0074$ &  $-$    \\[0.1cm]
    SMC742.26.330   & $0.345321$ & $1.68$ & $1.12$ & $350.85925$  & $-77.530417$ & $13.8640$ & $0.0042$   & $0.1020$ & $0.0049$ & $0.0074$ & $ 1.074$   \\[0.1cm]
    \hline
    \end{tabular}
    }
    \caption{Selected objects together with periods, estimated mass ratios, coordinates, Fourier coefficients ($A_i$) and parallaxes from Gaia DR3.}\label{objects}
\end{table}

Object GD1070.18.22288, contrary to other ones, was selected only due to the observed X-ray emission in the vicinity
that can indicate the existence of the accretion disk. However, it is much better fitted with the double star model than in the single star model,
X-ray emission was considered a much more important factor.
After the inclusion of GD1070.18.22288, fourteen objects in total were selected as potential dormant black hole candidates.
The key parameters of the selected objects are listed in Table \ref{objects}.
The parallax solution for LMC751.15.2886, initially considered statistically insignificant, was adopted in the end
as the ratio of parallax to statistical deviation was rather high (2.975) and
the RUWE parameter was close to unity (1.064), indicating that there was no problem with the Gaia astrometric solution. 
After this change, only SMC720.28.40576 is considered not to reside in the Galaxy.

What should be emphasized here is the meaning of the $\chi^2$/BIC score. Little is assumed about objects. Especially, the best fitting
$\log{g}$/Z is not investigated, which can be crucial to obtain a high-quality spectral fit. Due to this effect, it is unreasonable to compare the values of $\chi^2$/BIC
between objects, as few of them can be better fitted with the given metallicity/$\log{g}$
(for example, ultraviolet observations can be sensitive not only to the temperature of the star but also to its metallicity).
Even if some objects are not accurately described, it is still possible to reliably assess if one or more stars are contributing to the spectra.
Light curves for all objects (excluding SMC720.28.40576 and GD1070.18.2288) are presented in Appendix A while SED fits can be found in Appendix B.

\section{Results}
\subsection{Physical parameters of objects}
Each of the fourteen final objects was further investigated using available public data to determine the physical properties.
Of all objects, twelve are spectral type K, G, F, or A, one object is spectral type O, while the latter is composed of two stars.
Thirteen stars from the list have a measured parallax from Gaia DR3 while one of the objects is located in the SMC.

The mass of each object was estimated using Python package \texttt{isochrones} \citep{isochrones}. The package allows one to interpolate MIST tracks \citep{dotter_mist,Choi_mist}
assembled using MESA \citep{paxton_mesa_1,paxton_mesa_2,paxton_mesa_3}. In the case of objects with a statistically significant parallax
$\textrm{[Fe/H]} = 0$ was assumed, while in the case of the SMC $\textrm{[Fe/H]} = -1$ following \citet{metal_SMC}.
In order to find the best-fitting track $\chi^2=\frac{(\log L-\log L_{\textrm{SED}})^2}{(\Delta \log L) ^2}+\frac{(T-T_{\textrm{SED}})^2}{\Delta T ^2}$
value was minimized using the Nelder-Mead algorithm. Here, variables with the "SED" subscript stand for the estimates obtained from the SED fit, while values beginning with "$\Delta$"
denote associated errors obtained from MCMC chains. Infered parameters of the objects together with the MIST mass estimates are presented in Table \ref{objects_fit}.

\begin{table}%[H]
    \footnotesize
    \centerline{
    \centering
    \begin{tabular}{llllllll}
    \hline
    Name& $T$ [K] & $\log_{10} L/L_{\odot}$ &$E(B-V)$ & $d$ [kpc] & $\chi^2$ & N& $M_{MIST}$ [$\textrm{M}_\odot$]\\
    \hline
    \hline 
    \noalign{\vskip 2mm}    
    BLG931.27.36745 & $6454\pm 60$ & $0.896^{+0.044}_{-0.041}$ & $0.199$  & $0.69^{+0.03}_{-0.03}$ & $6.9$ & $19$ & $1.49$ \\[0.1cm]
    BLG986.08.7     & $6270^{+63}_{-65}$ & $0.823^{+0.015}_{-0.015}$ & $0.226$ & $0.83\pm 0.01$ & $7.5$ &   $17$ & $1.41$ \\[0.1cm]
    GD1097.20.23000 & $6040\pm 71$ & $0.636^{+0.013}_{-0.013}$ & $0.221$  & $0.56\pm 0.01$ & $11.7$ &$20$ & $1.27$ \\[0.1cm]
    GD1448.27.17    & $8261^{+483}_{-189}$ & $1.836\pm 0.023$ & $0.611$  & $1.30\pm 0.02$ & $8.8$ & $17$ & $2.52$ \\[0.1cm]
    GD2246.03.18414 & $7146^{+139}_{-117}$ & $0.855 \pm 0.016$ & $0.273$ & $0.71\pm 0.01$ & $58.4 $& $17$ & $1.50$ \\[0.1cm]
    LMC574.11.3407  & $4751^{+62}_{-58}$ & $-0.355^{+0.139}_{-0.123}$ & $0.034$  & $2.20^{+0.39}_{-0.29}$ & $2.4$ & $13$ & $-$ \\[0.1cm]
    LMC606.30.48    & $5565\pm 58$ & $-0.057\pm 0.064$ & $0.047$  & $1.93^{+0.15}_{-0.14}$ & $12.2$ & $15$ & $0.89$ \\[0.1cm]
    LMC751.15.2886  & $6237\pm 62$ & $0.472^{+0.330}_{-0.249}$ & $0.056$  & $5.54^{+2.73}_{-1.40}$ & $16.6$ & $15$ &  $1.16$ \\[0.1cm]
    MBR108.18.3     & $6264^{+37}_{-42}$ & $0.273\pm 0.014$ & $0.037$  & $0.91\pm 0.01$ & $37.8$ & $18$ & $1.18$ \\[0.1cm]
    MBR236.09.433   & $6750\pm 58$ & $0.801\pm 0.029$ & $0.068$  & $2.12\pm 0.07$ & $19.5$ & $18$ & $1.44$ \\[0.1cm]
    SMC711.22.1068  & $7037^{+47}_{-59}$ & $0.873 \pm 0.015$ & $0.021$  & $1.52\pm 0.03$ & $16.0$ & $19$ & $1.51$ \\[0.1cm]
    SMC720.28.40576 & $30169^{+2110}_{-1528}$ & $4.276 \pm 0.078$ & $0.095$  & $-$ & $28.0$ & $22$ & $12.4$ \\[0.1cm]
    SMC742.26.330   & $6263\pm 38$ & $0.422\pm 0.013$ & $0.083$  & $0.93\pm0.01$ & $21.8$ & $18$ & $1.18$ \\[0.1cm]
    \hline
    \end{tabular}
    }
    \caption{Estimated physical parameters of objects using the single star model together with $\chi^2$ values, number of observations $N$, 
    MIST mass estimates, and extinction estimates.}\label{objects_fit}
\end{table}

\subsection{Radial velocity semi-amplitude estimation from Gaia DR3 data}
Half of the objects in the final list have available high-quality radial velocity information that is normally computed for bright stars from the Gaia DR3 catalog \citep{gaia_collaboration_gaia_2023}.
While the estimate of the radial velocity is based on the median of measurements, 
the error of this estimate is based on the epoch standard deviation. According to \citet{katz_gaia_2023}, the radial velocity error $\delta v$ 
is calculated via the equation
\begin{align}
    \delta v=\sqrt{\sigma_{\textrm{med}}^2+0.11^2},\\
    \sigma_{\textrm{med}}=\sqrt{\frac{\pi}{2}}\frac{\sigma}{\sqrt{N}},
\end{align}
where $\sigma$ is a standard deviation of the radial velocity and $N$ stands for the number of transits used to compute radial velocity (all values are expressed in km\,s$^{-1}$).
This simple formula allows one to obtain the 
variance of radial velocity measurements as 
\begin{equation}
    \sigma^2=\frac{2N}{\pi}\left((\delta v)^2-0.11^2\right).
\end{equation}
It can be proven (for details see Appendix C) that if one assumes that the error is dominated by a sinusoidal radial movement with a semi-amplitude $K$,
a variance of velocity is equal to
\begin{equation}
    \sigma^2=\frac{K^2}{2}.
\end{equation}
This observation can be used to estimate a semi-amplitude of the radial velocity using Gaia measurements (denoted from here as $K_{\textrm{Gaia}}$) as 
\begin{equation}
    K_{\textrm{Gaia}}=\sqrt{\sigma^2}\sqrt{2}=2\sqrt{\frac{N}{\pi}\left((\delta v)^2-0.11^2\right)}.
\end{equation}
This radial velocity amplitude is then used to calculate a binary mass function, which provides a lower limit on the companions mass and can be obtained using the equation
\begin{equation}\label{mass}
    f(M_1,M_2,i)=\frac{M_2^3 \sin{i}^3}{(M_1+M_2)^2}=\frac{K^3 P}{2\pi G},
\end{equation}
where $P$ denotes an orbital period, while $G$ is the gravitational constant. For each object with a radial velocity estimate, a mass function was obtained
and it is presented in Table \ref{mass_function_table} with other relevant parameters.
For each object with a MIST mass estimate, a lower boundary of a companion mass was calculated by solving Eq. \ref{mass}
with $\sin{(i)}=1$ and placed in Table as $M_{\textrm{min}}$.

In \citet{katz_gaia_2023} a criterion was presented that allows one to test whether an object is variable in radial velocity.
The criterion states that objects with $N>10$, $T_{\textrm{eff}}\in [3900,8000]$\,K, {\it{rv\_chisq\_pvalue}} $<0.01$ and {\it{rv\_renormalised\_gof}} $>4$
can be safely considered to be variable in radial velocity. Each candidate was tested using the described method,
and only two entries (BLG931.27.36745 and GD1097.20.23000) cannot be safely assumed to be variable in radial velocity as they were not observed
enough times.

\begin{table}%[H]
    \footnotesize
    \centering
    \setlength{\tabcolsep}{3pt}
    \begin{center}
    \centerline{
    \begin{tabular}{llllllll}
    \hline
    Name & Period [d]& $RV$ [km/s]&$N$ & $\delta v$  [km/s]   & $K_{\textrm{Gaia}}$ [km/s]  & $f(M_1,M_2,i)$ [M$_{\odot}$]& $M_{\textrm{min}}$ [M$_{\odot}$]\\
    \hline
    \hline
    BLG931.27.36745 & $0.7050$   & $-4.35$ & $8$  & $14.81$ & $47.26$  & $0.0077$ & $0.29$\\[0.1cm]
    BLG986.08.7     & $0.5132$  & $-1.69$ & $15$ & $18.11$ & $79.14$   & $0.026$  & $0.45$\\[0.1cm]
    GD1070.18.22288 & $45.1467$  & $38.71$ & $20$ & $7.66$  & $38.65$  & $0.27$  & -- \\[0.1cm]
    GD1097.20.23000 & $0.4570$ & $36.5$  & $10$ & $21.03$ & $75.03$  & $0.020$  & $0.38$\\[0.1cm]
    GD1448.27.17    & $1.2410$  & $33.08$ & $22$ & $7.69$  & $40.69$  & $0.0087$ & $0.42$\\[0.1cm]
    GD2246.03.18414 & $0.4270$ & $27.68$ & $25$ & $23.03$ & $129.93$ & $0.097$  & $0.80$ \\[0.1cm]
    \hline
    \end{tabular}
    }
    \end{center}
    
    \caption{Estimated mass functions together with other relevant parameters.}\label{mass_function_table}
\end{table}

\subsection{Detailed analysis of objects}
\begin{figure}
   \includegraphics[width=1\textwidth]{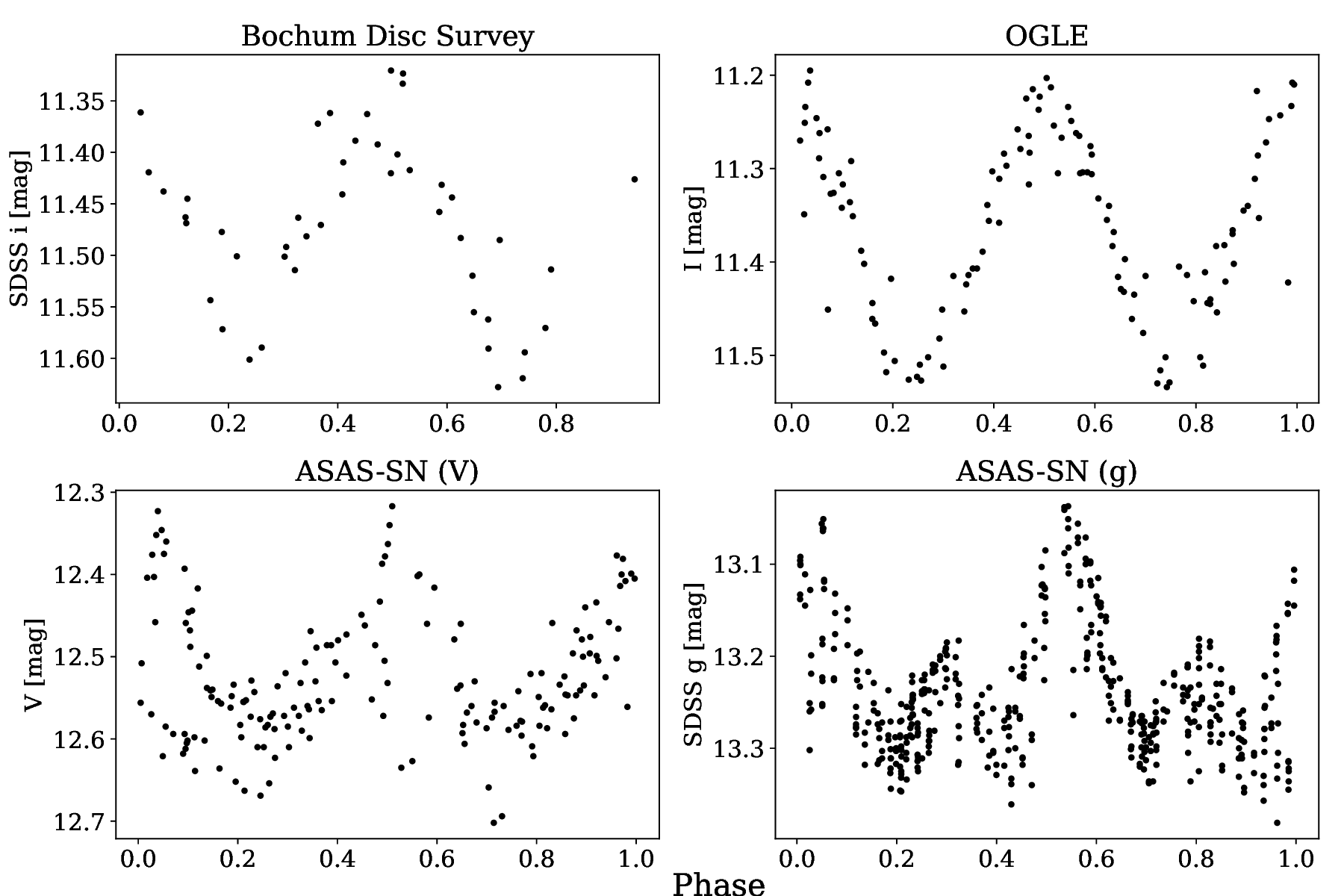}
   \caption{Light curve of GD1070.18.2288 in $4$ different bands folded with the OGLE period.}\label{comp}
\end{figure}

\begin{figure}
   \includegraphics[width=1\textwidth]{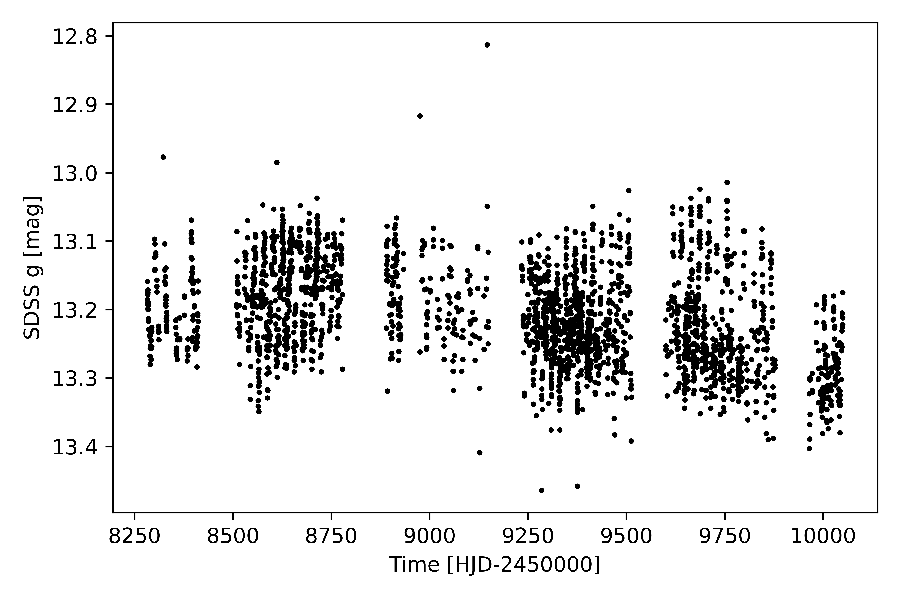}
   \caption{Change in brightness in the SDSS g band over time.}\label{evolution}
\end{figure}

In this section, a detailed description of the individual objects is provided. It is divided into three subsections, where the first two briefly cover the information
available for objects GD1070.18.22288 and SMC720.28.40576, while the third is dedicated to the remaining objects.
\paragraph{GD1070.18.22288}
\begin{figure}
    \includegraphics[width = \textwidth]{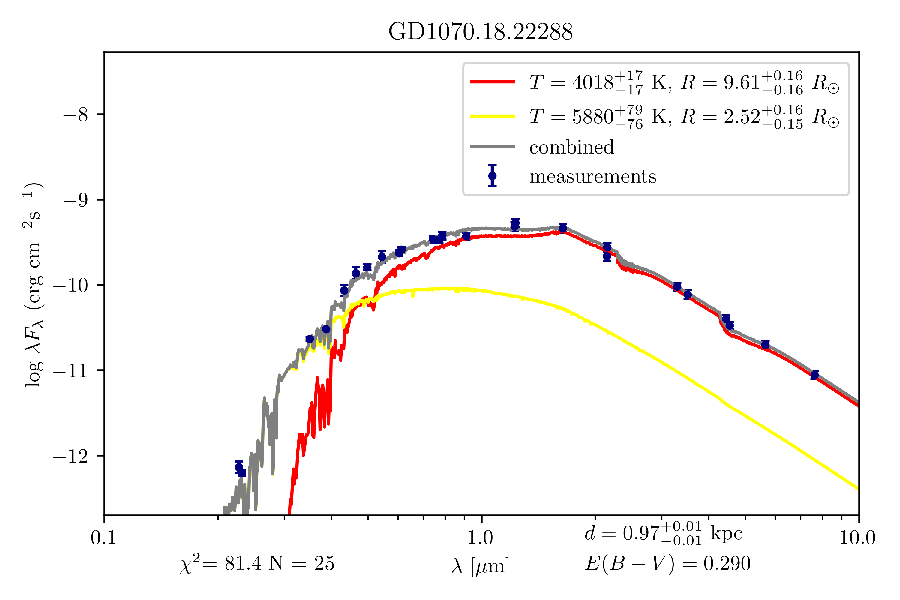}
    \caption{Spectral Energy Distribution fit for GD1070.18.22288 using a model with two components.}\label{GD1070SED}
\end{figure}
This object has the longest period from the sample $P\approx 45$\,d and it was listed as a plausible candidate only because of the reported X-ray emission.
Due to an excessive amount of observations in many parts of the spectra, it was possible to obtain a high-quality spectral energy distribution fit that revealed two sources
with different temperatures; detailed spectral distribution can be seen in Figure \ref{GD1070SED}.
One of the objects with a rather low effective temperature of
$T\approx 4000$\,K has radius $R\approx 10$ $\textrm{R}_{\odot}$ and it is quite far from any evolutionary track in the MIST database. This can indicate
that it has been stripped in the past by its companion.
The object can also be found in the ASAS-SN database \citep{jayasinghe_asas-sn_2019,sky_patrol,asas_sn} under the identification number J170801.81-410255.6 where it is classified
as a rotational variable star with half of the period found in this work. The ASAS-SN observed it in the $V$ and SDSS $g$ bands, allowing light curves to be compared with the OGLE curve.
Furthermore, the object was observed in the SDSS $i$ band by the Bochum Galactic Disk Survey \citep{hackstein_bochum_2015}. The comparison between all light curves can be found in Figure \ref{comp}.
ASAS-SN observations in the $g$ band can be traced back to $2016$ and allow one to gain a better understanding of the nature of the object.
The evolution of the light curve over time is presented in Figure \ref{evolution}.

GD1070.18.2288 was observed in the X-ray by Swift and XMM missions and published in serendipitous sources catalogs
(\citet{evans_2sxps_2020} and \citet{traulsen_xmm-newton_2020}).
The XMM X-ray source was observed a few times and is transient in nature, changing its flux slightly across time.
This area of the Galactic disk also came to attention because of the gamma-ray source HESS J1708-410 \citep{aharonian_hess_2008} located $4'.212$ from the estimated OGLE position. 
Gamma emission from the source has an extended character, and the position reported by the OGLE falls near the $1\sigma$ region of the emission centre.
To date no plausible explanation for the origin
of the aforementioned source was found, despite multiple efforts and searches in various parts of the optical spectra.
One particular work \citep{van_etten_multi-wavelength_2009} observed a close vicinity of
the gamma-ray source using the XMM Telescope and found faint X-ray emission from the point source labelled in the publication as nr $1$ (hereinafter referred to as [VFH2009] 1).
[VFH2009] 1, although quite faint, coincides with the position of GD1070.18.22280 with accuracy of $0''.4$.
In the publication it was found that X-ray emission is best fitted with an absorbed power law by hydrogen density $n_H=2.0\times 10^{21}$ $\textrm{cm}^{-2}$.
Interestingly, the assumed distance of $3$ kpc based on radio observations is nearly three times higher than the value obtained using the data from the Gaia DR3.
The X-ray luminosity based on the Gaia parallax roughly lies in the range $L_{X}=(1-3)\cdot10^{31}$ erg\,s$^{-1}$ depending on the time.\hfil \break%\par\hspace{\parindent}
\hspace*{17.62482 pt} As one can see in Figure \ref{comp} there is a discrepancy between the curves in the $I$ band and the curves from ASAS-SN.
This observation, together with the evolution of the amplitude of the variability as presented in Figure \ref{evolution},
suggests that the object represents a class of rotational variables.
This can also be partially supported by the X-ray emission from the system; it is widely known that many rotational variables like RS Canum Venaticorum can
emit X-rays with a luminosity around $\sim 10^{31}$ erg\,s$^{-1}$ \citep{walter_x-rays_1980}, so the value obtained from XMM 
is consistent with emission from this type of system. It is hard to determine whether the OGLE curve exhibits changes in brightness similar to the ASAS-SN light curve,
as observations cover only a relatively short period. 
It is highly unlikely that the variability in the $I$ band is caused by the ellipsoidal modulation.
If the period would be equal to $45$ d, the inferred radii would be too small for the system to fill their Roche lobes (unless the system would have a very small total mass).
The most plausible explanation states that the system is similar to the RS Canum Venaticorum variable, which has undergone mass transfer in the past.
The colder star is covered with spots that emerge due to high magnetic fields. Such stars can develop powerful coronal heating responsible for the X-ray emission.
One could investigate the X-ray spectrum of an object to find whether it is consistent with the emission from the hot plasma.
However, since the object does not represent a class of ellipsoidal variables, this line of investigation is dropped as it is beyond the 
scope of this work.

\begin{figure}%[H]
    \centering
    \includegraphics[width = \textwidth]{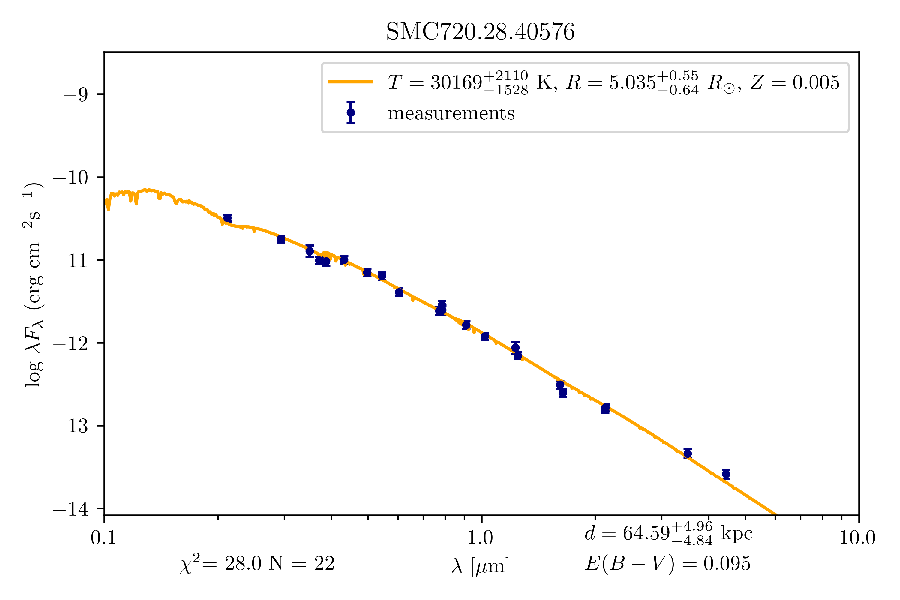}
    \caption{Single model SED fit for SMC720.28.40576.}
    \label{SMC720:sed}
\end{figure}

\paragraph{SMC720.28.40576}
\begin{figure}%[H]
      \centering
      \includegraphics[width = \textwidth]{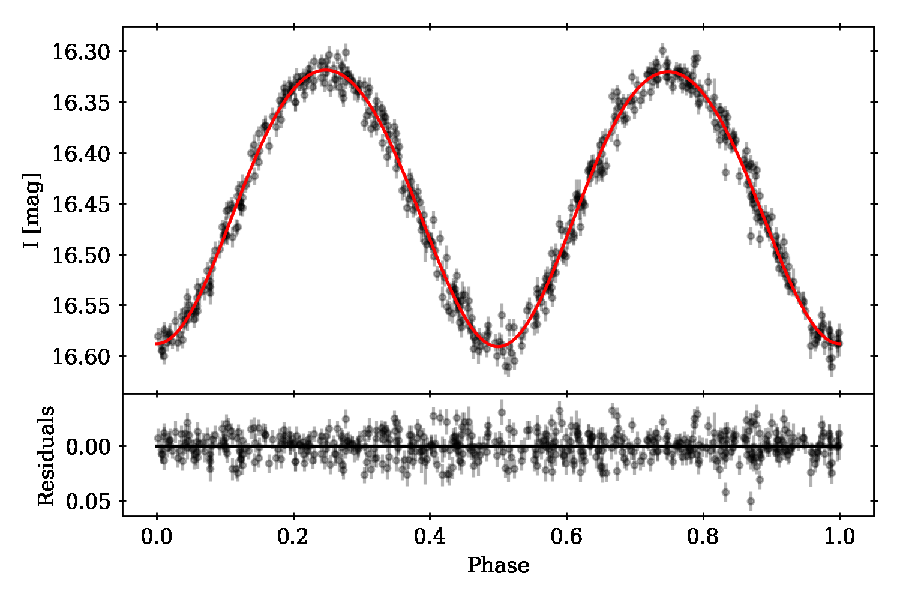}
      \caption{Light curve of SMC720.28.40576 collected by the OGLE project with 4th degree harmonic model.}
      \label{SMC720:lc}
\end{figure}
This object is the only one in the sample that is probably located in the Magellanic Clouds as the measured parallax is statistically insignificant (parallax to error ratio 
equal to $0.21$).
The estimated period of the variable is nearly half of the day suggesting rather a compact setup of stars.
The detailed light curve together with the residuals of the fourth-degree Fourier model is presented in Figure \ref{SMC720:lc}.
As the object lies in the region that was covered many times with X-ray observations to search for X-ray transients in the SMC,
a great number of measurements is available for the source. Especially, the optical monitor onboard the XMM mission measured the brightness of the source in the UV part of the spectrum 
allowing one to reveal the true nature of the object. 
The SED fit estimate suggests the star is of spectral type O. The photometric luminosity measured in each band together with the best-fitting spectra are presented in Figure \ref{SMC720:sed}.
The MIST evolutionary track estimates the mass of the object to be around $13$  $\textrm{M}_{\odot}$.
The star was already published in \citet{pawlak_ogle_2016} (under id name OGLE-SMC-ECL-1268) and classified as a contact binary. Although
no companion is visible in the SED distribution, it might hide itself in the light of a primary star.
No WISE source could be found, as the nearest detection is nearly $10''$ away from the position. The lack of any observations
in W3 and W4 bands is unfortunate as they would allow for the detection of cold companion (if present).

After assuming mass ratio equal to the $\tilde{q}_{\textrm{mmr}}$ from Table \ref{objects} and MIST estimate of the primary star's mass,
one can estimate that the companion should be more massive than $\sim 60$ M$_\odot$.
Such a great value of a companion's mass is quite puzzling, given the fact that it nearly reaches values from the high-mass gap.
On the other hand, such great amplitude of the light curve can indicate, that variability is caused by other modulation than ellipsoidal one.
Even if there is a main sequence star in the system (much less luminous then primary star) it would be really hard to explain the required mass of such an object. 

\paragraph{Remaining objects}
The majority of objects in the list can be characterized as objects with intermediate temperatures. They typically have a radius of
few solar radii and spectral types from K to A. Some objects from the Galactic disk sample can be found in ASAS-SN database \citep{jayasinghe_asas-sn_2019} as they are usually brighter 
on average than the objects in the Magellanic Clouds sample (as Gaia is able to compute radial velocity solution only for rather bright stars).
No detailed information has been found for the objects, although SMC742.26.330 was observed by the HERMES spectrograph
based on the Anglo-Australian telescope as part of the GALAH survey \citep{galah_main}. As the only one observation of the object was taken, no velocity information can be
obtained although very clear emission lines are present indicating ongoing mass transfer in the system. 
Temperature obtained from the GALAH spectrum is $T=6150\pm124$\,K validating the value obtained from the SED fit modelling.

\subsection{Contact binaries?}
\begin{figure}%{1\textwidth}
    \begin{adjustwidth}{-2cm}{-2cm}
    \centering
    \includegraphics[width=1.\textwidth]{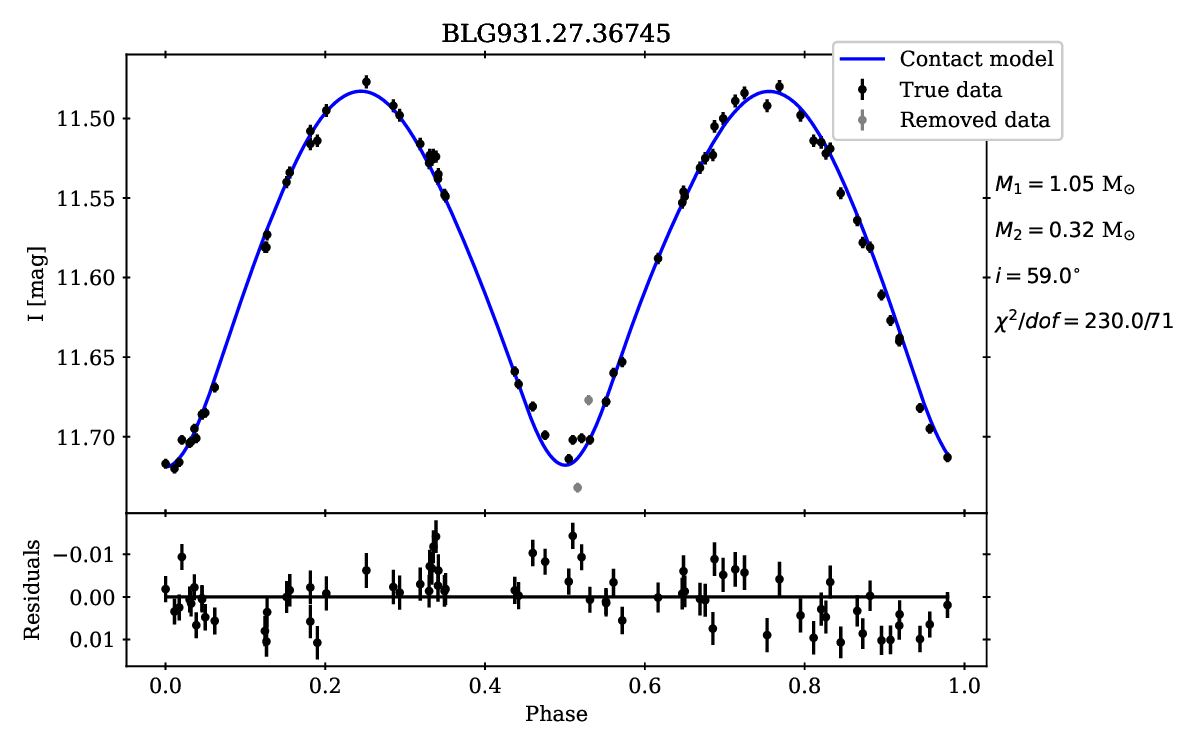}
    \caption{The light curve fit with the contact binary model for BLG931.27.36745.}\label{lc_plot}
    \end{adjustwidth}
\end{figure}

Although no detailed information about objects was found in the literature, two of them
are listed as nonsingle stars in the Gaia DR3 Eclipsing Binaries catalogue \citet{gaia_nonsingle} (BLG986.08.7, BLG931.27.36745).
The catalogue contains stars with collected light curves that matched the precomputed set of eclipsing/ellipsoidal binaries. 
Both objects reported by Gaia DR3 are fitted with the contact binary model where $f_1>1$ and $f_2>1$, with almost no
temperature difference (ratio close to $1$). This result points out that it is possible to obtain similar results
with the contact system of two similar stars. In such a compact setup, stars are in thermal equilibrium, and as a result,
the spectral energy distribution (SED) incorrectly fits them with a single-star model.
This line of inquiry was further investigated, and objects' light curves were fitted using the
PHOEBE simulation software \citep{wilson_realization_1971,prsa_computational_2005,conroy_physics_2020}.
The model was constructed in such a way that both stars shared the temperature $T$ (inferred from a single model SED) and were overflowing their Roche lobes.
Mainly objects from the Galactic disk sample were analyzed, as for them it is possible to compare estimates of radial velocity semi-amplitude with
values from the light-curve modelling. Additionally SMC720.28.40576 was analyzed in the similar setup as the object distinguishes itself from the 
rest of the sample with its remarkable companion's mass estimate.
Preprocessed light curves used to obtain the 4th degree harmonic fits were used (with sigma clipping outlier removal).
Light curves were fitted in the SED-like manner, for each phase value the magnitude in the $I$ band was predicted and compared with the value 
observed by OGLE. 
Two masses together with an inclination $i$ and a radius of the primary star $R_1$ form four parameters, which are fitted to minimize the $\chi^2$ value using the Nelder-Mead algorithm.
The second radius is no longer a free parameter, as it is controlled by a common envelope.
In the study, no flux normalization was performed. The extinction was set based on the values used to perform the spectral energy distribution fit, while distance 
estimates were based on photogeometric median distance calculated in \citet{distances_bailer}.
One could infer the effective luminosity of both stars from a SED fit although it should be noted, that the mean luminosity of the ellipsoidal binary is always smaller 
than the sum of the luminosity of both stars. Hence, in order to obtain a more realistic model, the total luminosity of the system (parametrized by the radius of a primary star) can be 
changed. To calculate the radial velocity semi-amplitude of the primary star one can simply rewrite Eq. (\ref{mass}) to obtain:
\begin{equation}
    K_{\textrm{estimate}}=M_2\sin{(i)}\left(\frac{2\pi G}{P(M_1+M_2)^2}\right)^{\frac{1}{3}}.
\end{equation}
The fitted values together with the $\chi^2$ values are presented in Table \ref{light_curve_fits}
together with the predicted semi-amplitude of the velocity and the estimated one from Gaia DR3 data.
An example light curve fit for the object BLG931.27.36745 can be seen in Figure \ref{lc_plot}.

\begin{table}%[H]
    \footnotesize
    \begin{center}
    \centerline{
    \begin{tabular}{llllllllll}
    \hline
    Name & $M_1$ [M$_{\odot}$] & $M_2$ [M$_{\odot}$]&$R_1$ [$R_{\odot}$] & $R_2$ [$R_{\odot}$] & $i$ & $K_{\textrm{estimate}}$ [km/s] & $K_{\textrm{Gaia}}$ [km/s]&$\chi^2/dof$ \\
    \hline
    \hline
    BLG931.27.36745 & $1.05$ & $0.32$ &$1.86$ & $1.11$& $58.9^\circ$ & $53.0$ & $47.3$ & $230.0/71$ \\[0.1cm]
    BLG986.08.7 & $1.57$ & $0.40$& $1.76$ & $0.98$ & $56.0^\circ$ & $55.6$ & $79.1$ & $105.6/70$ \\[0.1cm]
    GD1097.20.2300 & $1.33$ & $1.17$ & $1.39$& $1.31$ & $60.68^\circ$ &$152.5$& $75.0$ & $167.9/93$ \\[0.1cm]
    GD1448.27.17 & $1.65$ & $0.79$ & $3.16$& $2.34$ & $51.2^\circ$ & $67.4$ & $40.7$ & $854.1/136$\\[0.1cm]
    GD2246.03.1814 & $1.18$ & $1.04$& $1.26$& $1.03$ & $58.3^\circ$ & $119.5$ & $129.9$ & $510.9/101$\\[0.1cm]
    \hline
    \end{tabular}
    }
    \caption{Parameters used to fit the light curves of objects together with the radius of the secondary star, the estimates of radial velocities, and the normalized $\chi^2$ values.}\label{light_curve_fits}
    \end{center}
\end{table}
All light curves are presented in Appendix D.
Out of five objects, two can be well explained by the contact model with an intermediate mass ratio that is typical for this type of binaries.
Mass transfer, magnetic breaking, and other physical phenomena may lead to unequal masses resulting in extreme mass ratio binaries. For two other objects
a rather high mass ratio is preferred, which is unusual for such binaries. This is supported in the case of GD2246.03.1814 by the radial velocity estimate from Gaia DR3,
while it is not supported in the case of GD1097.20.2300. 
Although such a high mass fraction is quite uncommon among contact binaries, it 
should be emphasized that during the preselection process, only sources with high amplitudes of radial velocity from Gaia were selected.
Hence, the final sample can be nonrepresentative as this strict criterium introduces the selection effect. 
The last object in the list (GD1448.27.17) is not fitted well with the contact model. This can be
an indicator of spot variability that causes nonequal maxima. 

\begin{figure}%{1\textwidth}
    \begin{adjustwidth}{-2cm}{-2cm}
    \centering
    \includegraphics[width=1.\textwidth]{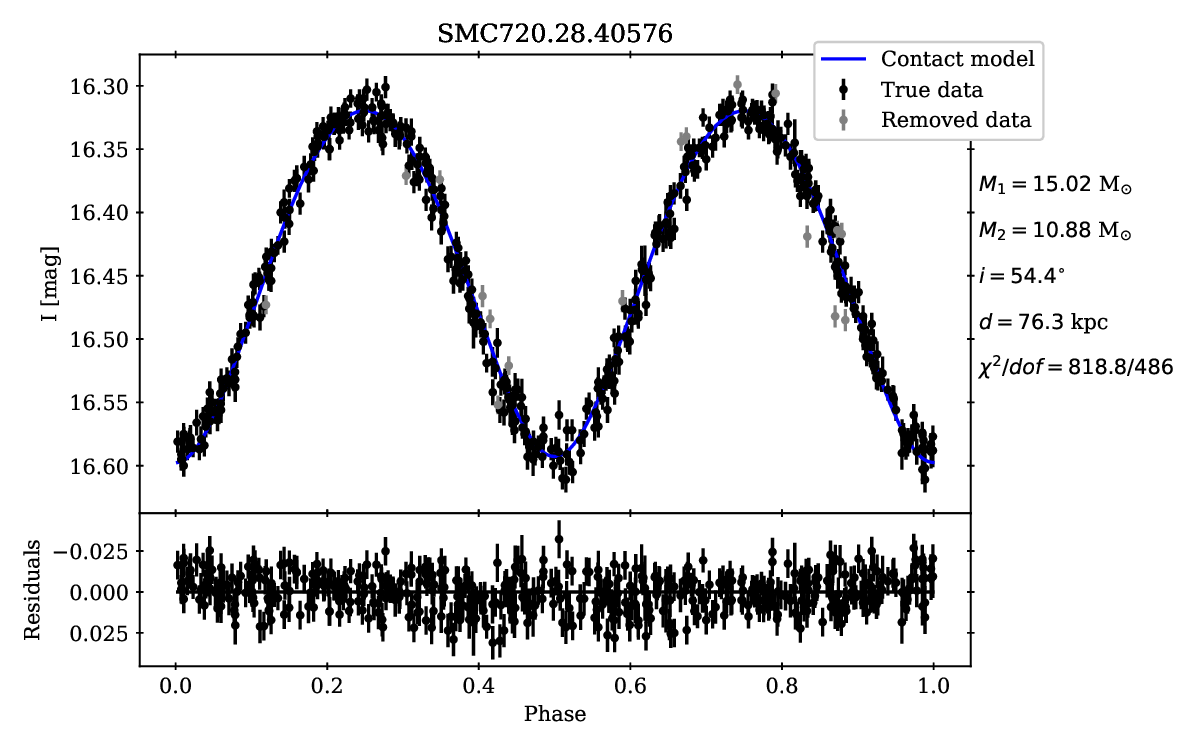}
    \caption{The light curve fit with the contact binary model for SMC720.28.40576.}\label{lc_fit_SMC720}
    \end{adjustwidth}
\end{figure}

Similarly to the candidates from the Galactic disk sample, SMC720.28.40576 was analyzed with an analogous setup.
Unfortunately, hot stars need other treatment in PHOEBE so the blackbody atmosphere was used with a bolometric
gravity brightening coefficient of $1$, a reflection coefficient of
$1$ and a logarithmic limb darkening law.
As the object lies in the SMC, no detailed distance is known. It was decided that the distance would also be considered
a free parameter. Additional loss $\frac{(d-d_{SMC})^2}{\sigma_d^2}$ was introduced to penalize the model for huge distances.
Using the Nelder-Mead algorithm it was found that the light curve can be well explained using a $15 M_\odot + 10.9 M_\odot$ model with the radii of primary and secondary equal to
$4.06 R_\odot$ and $3.60 R_\odot$ respectively, the distance $76.3$ kpc and the inclination  of $54.4^\circ$. The light curve with the fitted contact model is presented in Figure \ref{lc_fit_SMC720}.
Not only the binary is contact in nature, but respective volume fillout factors defined as the ratio of the stellar volume divided by the roche lobe volume are
$f_{V1} = 1.59 $ and $f_{V2} = 1.72$ suggesting a high degree of connectivity. 
Moreover, one can see that the system is exceptionally massive; up to this day only ten O+O overconntact massive binaries have been found \citep{overcontact-O}.
If the contact hypothesis were true, SMC720.28.40576 would have one of the shortest periods among all known massive overcontact binaries to date.

The values obtained from the fit can be quite distant from reality, as there is not enough data to constrain the parameters governing the model
(especially since no detailed radial velocity information is available).
Despite this fact, as demonstrated, not only some objects can be well described using the contact model, but in all cases, the contact model can achieve
an amplitude comparable to the observed one. Hence, contamination from such binaries is one of the most significant weaknesses of the method described
in \citet{gomel_search_2021a,gomel_search_2021b,gomel_search_2021c} as a simple contact binary can easily be classified as a potential candidate for 
dormant black hole system.

\section{Discussion \& Conclusions}

Of the total $8515$ objects investigated in this study, only fourteen were classified as plausible candidates for compact companion binaries.
Only one object in the sample is probably located in the Magellanic Clouds, while the rest of them reside inside the Milky Way. 
This is not unexpected as after limiting $I<17$ mag, only the most luminous objects inside the Magellanic Clouds can pass this threshold.
Further investigation into the sample revealed one rotational variable and suggested an alternative explanation for the remaining objects in the form of
contact binaries with intermediate mass ratios. It is impossible to pinpoint the exact nature of objects as only spectroscopic observations allow one to determine the 
radial velocity of stars and the true nature of companion objects. All of those candidates can potentially host neutron stars/black holes, but
due to the great amounts of contact binaries in the universe, contact scenarios seem to be much more plausible.
Gaia DR3 semi-amplitude estimations suggest that most of the companions are not compact at all. Even for stars with the estimated radial velocity semi-amplitude of $\sim 50$ km/s,
the companion's mass lower limit is too low for a compact star.
Only one object distinguishes itself from the rest with the high estimate of the velocity semi-amplitude $\approx120$ km/s. 
Based on the MIST estimate, a lower limit for companion mass is equal to $\approx0.8$ M$_{\odot}$ making the candidate quite promising. 

Until now, only one publication tried to perform any kind of follow-up observations of candidates found with the method described in \citet{gomel_search_2021a,gomel_search_2021b,gomel_search_2021c}.
In \citet{nagarajan_spectroscopic_2023} a sample of objects from the list presented in \citet{gomel_gaia_2022} was selected
and a time series of radial velocities was obtained. It was found that all selected objects cannot host any type of compact object as the velocity semi-amplitudes are too low.
It was suggested that such variables can be easily explained by contact binaries with or without spots. This conclusion is similar in some sense to the results presented in this work.
Such an explanation is quite convincing as such objects should be quite common, contrary to the black-hole binaries. It was pointed out that binary stars with the nearly overflowing
primary star should be rather rare as this phase of a binary evolution is very short-lived. Hence, the system should be either well-detached or overflowing its Roche lobe,
resulting in an X-ray binary.  In fact in \citet{gomel_search_2021a,gomel_search_2021b,gomel_search_2021c} the presented method was applied to some X-ray binaries with black hole companions and
it was found that at least some of them would be detected.
As noted in the introduction, the creation of such systems is rather rare, so the potential sample of black holes that can be revealed seems
to be small compared to black holes in wider binaries.
Moreover, as demonstrated, a simple contact binary can be easily classified as a candidate black hole. 
Therefore, the potential sample can be primarily populated by those common false positive contaminants.
Although \citet{nagarajan_spectroscopic_2023} did not find any compact companion,
more objects should be investigated to better characterize the population of these high-amplitude binaries.

\Acknow{
We thank all the OGLE observers for their contribution
to the collection of the photometric data over the decades.

We would like to thank Milena Ratajczak for the discussion.
Work by MK and PM was supported by the grant OPUS/2021/41/B/ST9/00252 of National Science Center, Poland awarded to PM.
For the purpose of Open Access, the author has applied a CC-BY public copyright license to any Author Accepted Manuscript (AAM) version arising from this submission.

This work has made use of data from the European Space Agency (ESA) mission
{\it Gaia} (\href{https://www.cosmos.esa.int/gaia}{https://www.cosmos.esa.int/gaia}), processed by the {\it Gaia}
Data Processing and Analysis Consortium (DPAC, \href{https://www.cosmos.esa.int/web/gaia/dpac/consortium}{https://www.cosmos.esa.int/web/gaia/dpac/\newline consortium}). Funding for the DPAC
has been provided by national institutions, in particular the institutions
participating in the {\it Gaia} Multilateral Agreement.
}
\bibliographystyle{acta}
\bibliography{bibliography}

\begin{NoHyper}
\appendix

\section{Light Curves}

\begin{figure}[H]
    \begin{adjustwidth}{-2cm}{-2cm}
    \centering
    \includegraphics[width=1.4\textwidth]{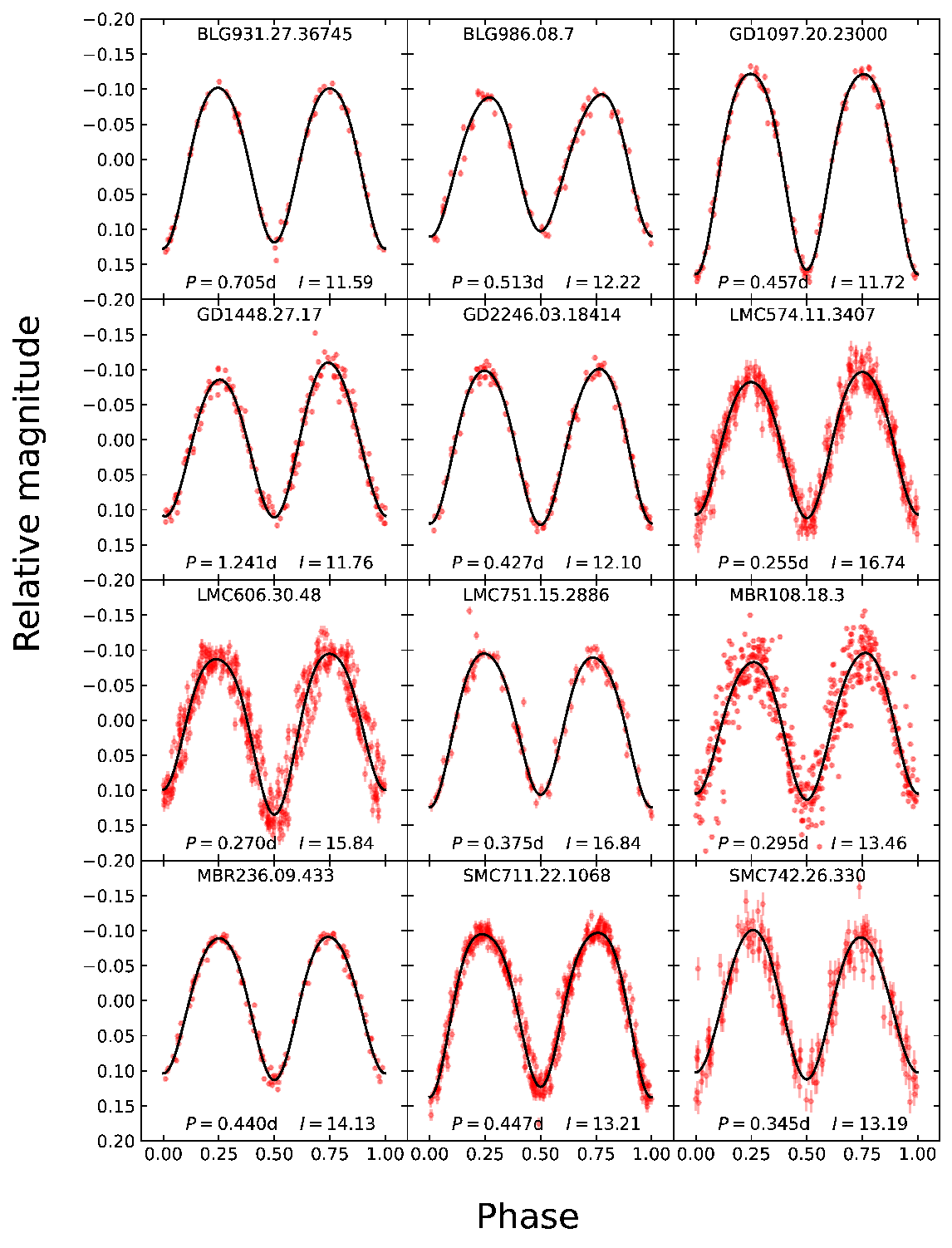}
    \end{adjustwidth}
\end{figure}

\section{Spectral Energy Distribution plots}

\begin{figure}[H]
    \begin{adjustwidth}{-2cm}{-2cm}
    \centering
    \includegraphics[width=1.4\textwidth]{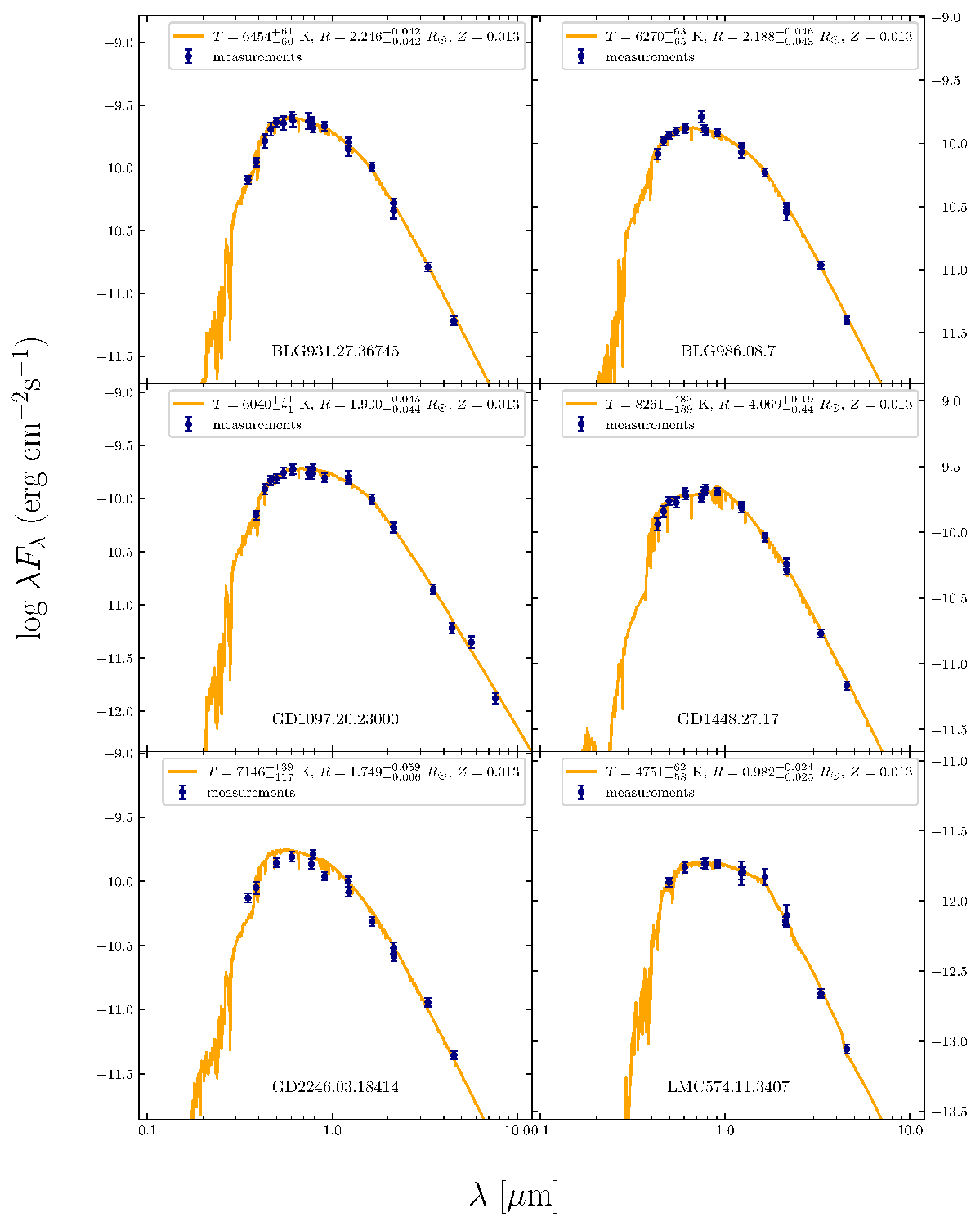}
    \end{adjustwidth}
\end{figure}
\begin{figure}[H]
    \begin{adjustwidth}{-2cm}{-2cm}
    \centering
    \includegraphics[width=1.4\textwidth]{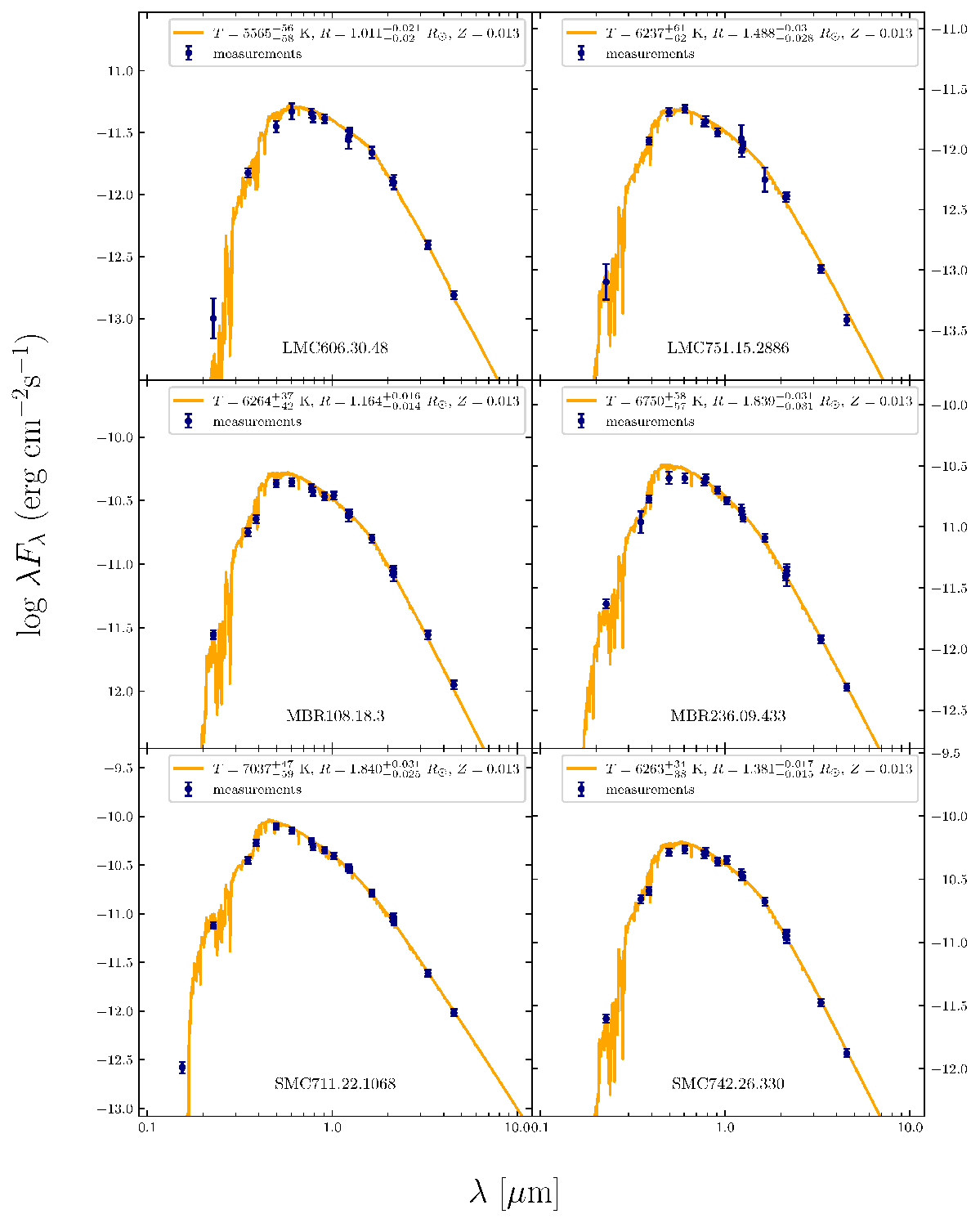}
    \end{adjustwidth}
\end{figure}
\let\cleardoublepage\clearpage
\section{Semi-amplitude estimate from Gaia DR3}
Let us assume that a radial velocity can be factored into two separate movements: the centre of mass movement (with velocity $v_0$)
and the circular motion with semi-amplitude $K$. Under the following assumptions, one can find that samples of radial velocity (denoted as $V_i$) can be written as 
\begin{equation}
    V_i=v_0+K\cos{(2\pi X_i)}
\end{equation}
where $X_i$ denote a sample from $\mathcal{U}(0,1)$.
Now, one can calculate the variance of this variable using identity $Var(x)=\mathbb{E} X^2 -(\mathbb{E}X)^2$.
\begin{align}
    \begin{split}
    \mathbb{E} V &= V_0 + \mathbb{E }K\cos{(2\pi X)} = V_0\\
    \mathbb{E} V^2 &= \mathbb{E} \left(V_0^2+\frac{K^2}{2}(1-\cos{(4\pi X)})+2V_0K\cos{(2\pi X)}\right)\\
                 &= V_0^2+\frac{K^2}{2}
    \end{split}
\end{align}
using the fact, that the expected value of the sinusoidal signal is zero.
Hence 
\begin{equation}
    Var(V) = \frac{K^2}{2}+V_0^2-V_0^2=\frac{K^2}{2}.
\end{equation}
\newpage

\section{Light curve fits}
\begin{figure}[H]
    \begin{adjustwidth}{-2cm}{-2cm}
    \begin{subfigure}{\textwidth}
        \includegraphics[width = 1.4\textwidth]{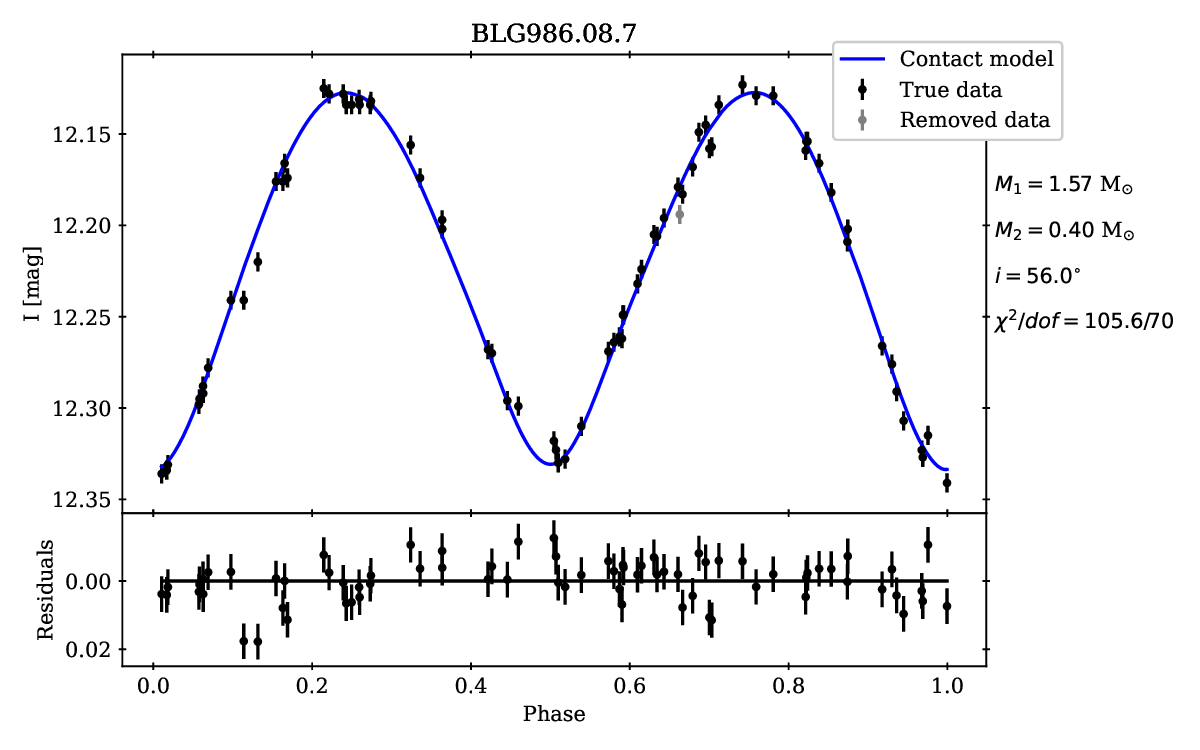}
    \end{subfigure}
    \begin{subfigure}{\textwidth}
        \includegraphics[width = 1.4\textwidth]{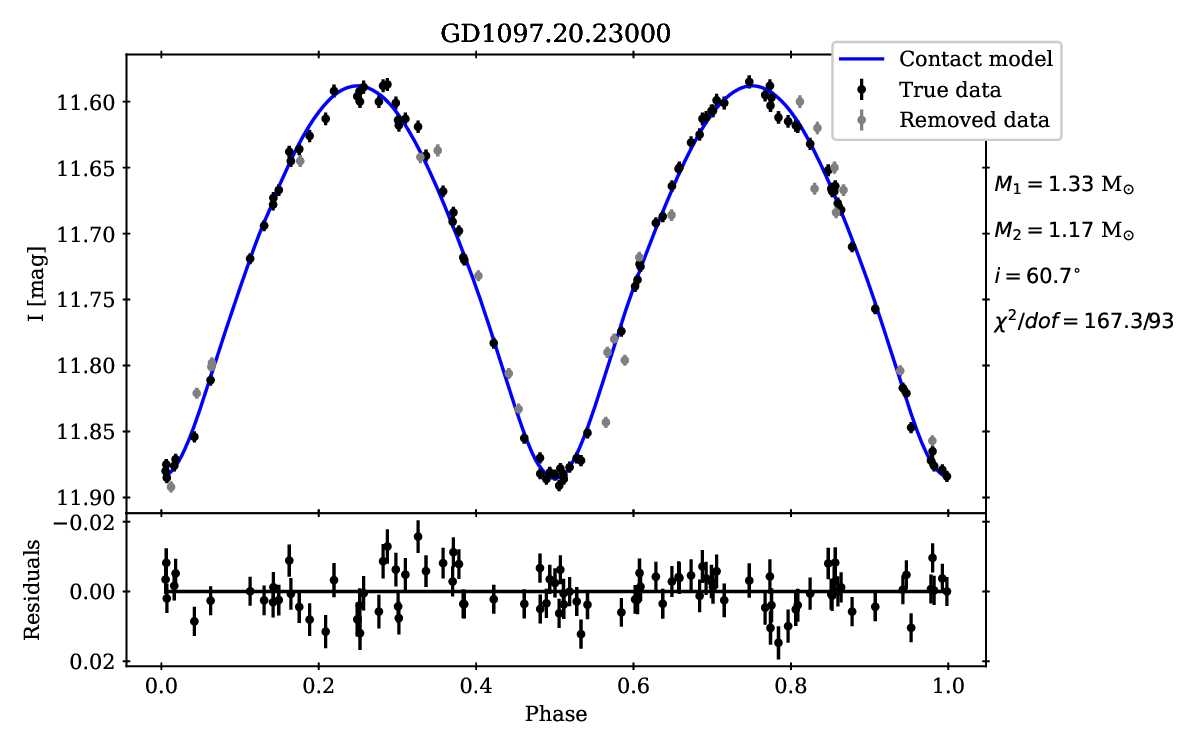}
    \end{subfigure}
    \end{adjustwidth}
\end{figure}
\begin{figure}[H]
    \begin{adjustwidth}{-2cm}{-2cm}
    \begin{subfigure}{\textwidth}
        \includegraphics[width = 1.4\textwidth]{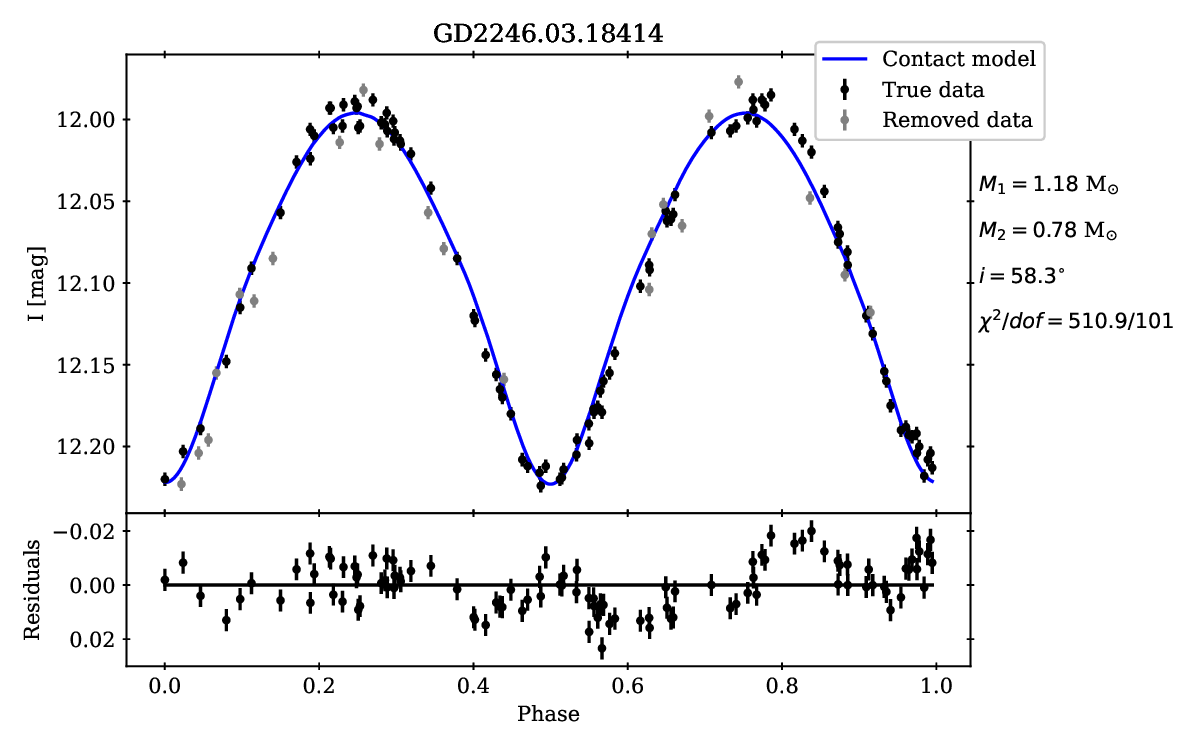}
    \end{subfigure}
    \begin{subfigure}{\textwidth}
        \includegraphics[width = 1.4\textwidth]{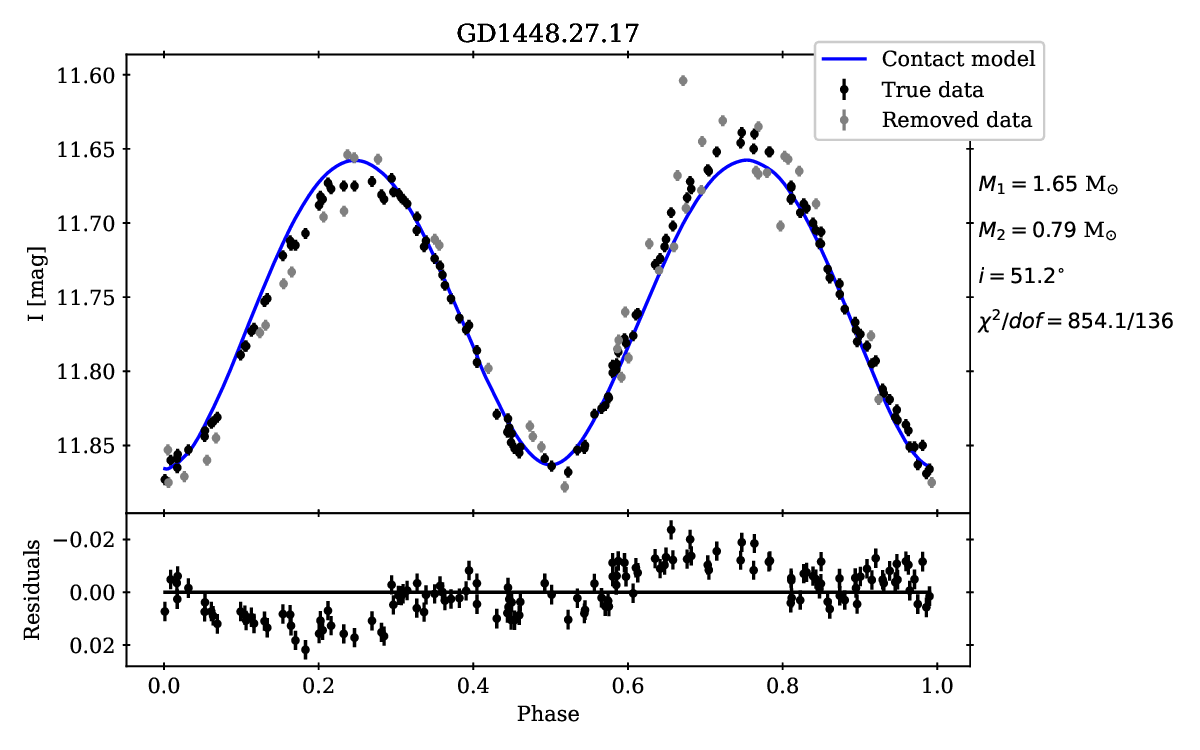}
    \end{subfigure}
    \end{adjustwidth}
\end{figure}
\end{NoHyper}
\end{document}